\documentclass[12pt]{article}
\usepackage{latexsym,amssymb,amsfonts,amsmath}
\usepackage[dvips]{graphicx}
\usepackage[usenames]{color}

\newlength{\extraspace}
\newlength{\extraspaces}
\newcommand{\be}{\begin{equation}
\addtolength{\abovedisplayskip}{\extraspaces}
\addtolength{\belowdisplayskip}{\extraspaces}
\addtolength{\abovedisplayshortskip}{\extraspace}
\addtolength{\belowdisplayshortskip}{\extraspace}}
\newcommand{\ee}{\end{equation}}

\newcommand{\bea}{\begin{eqnarray}
\addtolength{\abovedisplayskip}{\extraspaces}
\addtolength{\belowdisplayskip}{\extraspaces}
\addtolength{\abovedisplayshortskip}{\extraspace}
\addtolength{\belowdisplayshortskip}{\extraspace}}
\newcommand{\eea}{\end{eqnarray}}

\newcommand{\newsection}[1]{
\vspace{15mm}
\pagebreak[3]
\addtocounter{section}{1}
\setcounter{equation}{0}
\setcounter{subsection}{0}
\setcounter{footnote}{0}
\begin{flushleft}
{\Large\bf \thesection. #1}
\end{flushleft}
\nopagebreak
\medskip
\nopagebreak}

\def\Tr{{\rm Tr}}
\def\re{{\rm Re}}
\def\im{{\rm Im}}

\definecolor{mygray}{gray}{0.5}
\definecolor{mygray2}{gray}{0.3}

\begin{document}

\addtolength{\baselineskip}{.8mm}

{\thispagestyle{empty}

\noindent \hspace{1cm}  \hfill Revised version \hspace{1cm}\\
\mbox{}                 \hfill October 2008 \hspace{1cm}\\

\begin{center}
\vspace*{1.0cm}
{\large\bf HIGH--ENERGY HADRON--HADRON (DIPOLE--DIPOLE) SCATTERING
FROM LATTICE QCD}
\\
\vspace*{1.0cm}
{\large Matteo Giordano\footnote{E--mail: matteo.giordano@df.unipi.it}
  and Enrico Meggiolaro\footnote{E--mail: enrico.meggiolaro@df.unipi.it} }\\
\vspace*{0.5cm}{\normalsize
{Dipartimento di Fisica, Universit\`a di Pisa,
and INFN, Sezione di Pisa,\\
Largo Pontecorvo 3,
I--56127 Pisa, Italy.}}\\
\vspace*{2cm}{\large \bf Abstract}
\end{center}

\noindent
In this paper the problem of high--energy hadron--hadron (dipole--dipole)
scattering is approached (for the first time) from the point of view
of lattice QCD, by means of Monte Carlo numerical simulations.
In the first part, we give a brief review of how
high--energy scattering amplitudes can be reconstructed, using a
functional--integral approach, in terms of certain correlation
functions of two Wilson loops and we also briefly recall some relevant
analyticity and crossing--symmetry properties of these loop--loop
correlation functions, when going from Euclidean to Minkowskian theory.
In the second part, we shall see how these (Euclidean) loop--loop correlation
functions can be evaluated in lattice QCD and we shall compare our numerical
results with some nonperturbative analytical estimates that appeared in the
literature, discussing in particular the question of the analytic continuation
from Euclidean to Minkowskian theory and its relation to the still unsolved
problem of the asymptotic $s$--dependence of the hadron--hadron total cross
sections.
\\
\\
}
\newpage

\newsection{Introduction}

One of the most important open problems in hadronic physics (studied since
long before the discovery of QCD) is to explain/predict the (asymptotic)
high--energy behaviour of hadron--hadron total cross sections.
Present--day experimental observations (up to center--of--mass total energy
$\sqrt{s} = 1.8$ TeV) seem to be well described by a {\it pomeron}--like
high--energy behaviour (see, for example, Ref.~\cite{pomeron-book} and
references therein):
\be
\sigma_{\rm tot}^{(hh)} (s) \mathop{\sim}_{s \to \infty}
\sigma_0^{(hh)} \left( {\dfrac{s}{s_0}} \right)^{\epsilon_P} ,
~~~~{\rm with}~~\epsilon_P \simeq 0.08 .
\label{pomeron}
\ee
This behaviour is known in the literature as the {\it soft pomeron},
to be distinguished from the well--known BFKL (or {\it hard}) {\it
  pomeron}~\cite{BFKL}, i.e., $\sim s^{{\frac{12\alpha_s}{\pi}}\log 2}$, 
with $\alpha_s = g^2/4\pi$, obtained in perturbative QCD.
As we believe QCD to be the fundamental theory of strong interactions,
we also expect that it correctly predicts from first principles the behaviour
of hadronic total cross sections with energy. Anyway, in spite of all the
efforts, a satisfactory solution to this problem is still lacking.\\
We should also remind the reader at this point that the {\it pomeron}--like
behaviour (\ref{pomeron}) is, strictly speaking, theoretically forbidden (at
least if considered as a true {\it asymptotic} behaviour) by the well--known
Froissart--Lukaszuk--Martin (FLM) theorem~\cite{FLM} (see
also~\cite{Heisenberg}), according to which, for $s \to \infty$, 
$\sigma_{\rm tot}(s) \le {\frac{\pi}{m_\pi^2}} \log^2 \left( {\frac{s}{s_0}}
\right)$,
where $m_\pi$ is the pion mass and $s_0$ is an unspecified squared mass scale.
In this respect, the {\it pomeron}--like behaviour (\ref{pomeron}) can at most
be regarded as a sort of {\it pre--asymptotic} (but not really
{\it asymptotic}!) behaviour of the high--energy total cross sections
(see, e.g., Refs.~\cite{BB,Balitsky,Kaidalov} and references therein), valid
in a certain high--energy range.

From a general theoretical point of view, the {\it optical theorem} (which is
a consequence of unitarity) allows one to derive hadron--hadron total
cross sections from the expressions of the hadron--hadron {\it elastic}
scattering amplitudes ${\cal M}_{(hh)} (s, t)$ ($t$ being the transferred
momentum squared):
\be
\sigma_{\rm tot}^{(hh)} (s) \mathop{\sim}_{s \to \infty}
{\dfrac{1}{s}} {\rm Im} {\cal M}_{(hh)} (s, t=0).
\label{optical}
\ee
High--energy hadron--hadron elastic scattering processes at low
transferred momentum (usually called {\it soft high--energy
  scattering} processes) possess two different and widely separated
energy scales: the center--of--mass total energy squared $s$, which is
a hard scale ($s \gg 1$ GeV$^2$: formally we consider the limit
$s\to\infty$), and the transferred momentum squared $t$, which is a
{\it soft} scale, smaller than (or nearly equal to) the typical
energy scale of strong interactions ($|t| \lesssim 1~ {\rm GeV}^2 \ll
s$).  In this energy regime we cannot fully rely on perturbation
theory and a genuine nonperturbative treatment is in order.

Since Nachtmann's seminal paper in 1991~\cite{Nachtmann91} a lot of
work has been done on the problem of {\it soft high--energy
  scattering} in the framework of nonperturbative
QCD~\cite{DFK,Nachtmann97,BN,Dosch,LLCM1}.  
Using a functional--integral approach, hadron--hadron elastic
scattering amplitudes are reconstructed from the correlation functions
of certain Wilson loops, defined in Minkowski space--time (see Section
2 below). As it has been shown
in~\cite{Meggiolaro97,Meggiolaro98,Meggiolaro02,Meggiolaro05}  
(see Section 2 below), such correlation functions can be recovered
after proper analytic continuation from their Euclidean counterparts,
i.e., correlation functions of certain Wilson loops forming an angle
$\theta$ in Euclidean space: this has paved the way to the application
of nonperturbative techniques, which are normally available only in
the Euclidean formulation of QCD. Some calculations of the loop--loop
Euclidean correlation functions already exist in the literature (see
Section 4 below), using the so--called {\it Stochastic Vacuum Model}
(SVM)~\cite{LLCM2}, the {\it Instanton Liquid Model}
(ILM)~\cite{instanton1}, or the AdS/CFT correspondence for strongly
coupled gauge theories in the limit of a large number of
colours~\cite{JP1,JP2,Janik}. The Euclidean correlator is then
continued to the corresponding Minkowskian correlation function using
the above--mentioned analytic continuation in the angular variables.
The loop--loop correlation functions (both in the Minkowskian and in the
Euclidean theories) have also been computed exactly in the first two
orders of perturbation theory, ${\cal O}(g^4)$ and ${\cal O}(g^6)$, in
Ref.~\cite{BB}. 

Of course, each of these models, used in the calculation of the Euclidean
correlation functions, has its own {\it limitations}, which are
reflected in the variety of answers in the literature: someone finds
constant cross sections, someone else finds a {\it soft--pomeron}
behaviour, someone else finds a {\it hard--pomeron} behaviour $\ldots$
(And maybe the true asymptotic behaviour is $\log^2(s/s_0)$, thus
saturating the FLM bound!?). Unfortunately, these limitations are
often out of control, in the sense that one does not know how much
information is lost due to the involved approximations.
This is surely a crucial point which, in our opinion, should
be further investigated. (For example, in Ref.~\cite{Meggiolaro07} it has
been shown that the Euclidean--to--Minkowskian
analytic--continuation approach can, with the inclusion of some extra,
more or less plausible, {\it assumptions}, easily reproduce a
pre--asymptotic {\it pomeron}--like behaviour,
like the one in (\ref{pomeron}), which violates the FLM bound.)

A real breakthrough in this direction could be provided by a direct numerical
calculation of the loop--loop Euclidean correlation function in Lattice
Gauge Theory (LGT). In particular, one is interested in the dependence of the
correlation function on the angle $\theta$ between the loops, from
which the energy dependence of the physical scattering amplitudes can be
derived after a proper analytic continuation to Minkowski space--time.
Clearly a lattice approach can at most give the above--mentioned
function only for a finite set of $\theta$--values, from which it is clearly
impossible (without some extra assumption on the interpolating continuous
function) to get, by analytic continuation, the corresponding Minkowskian
correlation function (and, from this, the elastic scattering amplitudes and
the total cross sections).
However, the lattice approach, which provides a real first--principle
determination of the loop--loop correlator in Euclidean space, can be used
to investigate the goodness of a given existing analytic model (such as
SVM, ILM, AdS/CFT, and so on $\ldots$) or even to open the way
to some new model, simply by trying to fit the lattice data with the
considered model. This is exactly what we shall try to do in this paper.

The plan of this paper is the following.
In Section 2 we give (for the benefit of the reader) a brief review of how
high--energy scattering amplitudes can be reconstructed, using a
functional--integral approach, in terms of certain correlation
functions of two Wilson loops and we also briefly recall some relevant
analyticity and crossing--symmetry properties of these loop--loop
correlation functions, when going from Euclidean to Minkowskian theory.
In Section 3 we shall see how these (Euclidean) loop--loop correlation
functions can be evaluated in lattice QCD and discuss the main technical
complications.
In Section 4 we shall compare our numerical results with some
analytical nonperturbative estimates that appeared in the literature,
discussing in particular the question of the analytic continuation 
from Euclidean to Minkowskian theory and its relation to the still unsolved
problem of the asymptotic $s$--dependence of the hadron--hadron total cross
sections.
In Section 5 we draw our conclusions and show some prospects for the future.

\newsection{Hadron--hadron (dipole--dipole) scattering amplitudes from
Wilson loop correlators}

\begin{figure}[htb]
\includegraphics{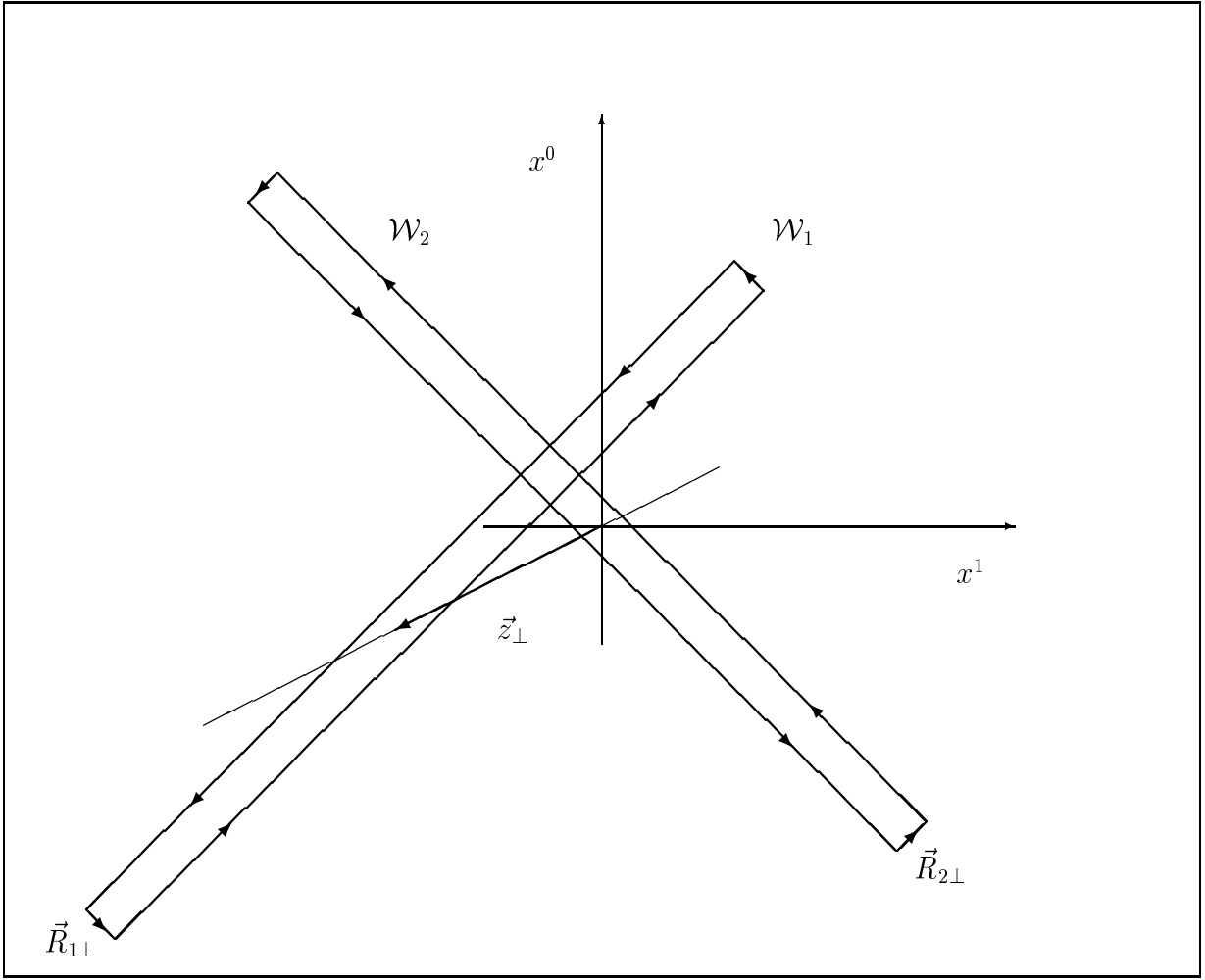}
\vspace{1.5cm}
\caption{The space--time configuration of the two Wilson
loops ${\cal W}_1$ and ${\cal W}_2$ entering in the expression for
the dipole--dipole elastic scattering amplitude in the high--energy limit.}
\label{fig:1}
\end{figure}

\noindent
It was shown in Refs.~\cite{DFK,Nachtmann97,BN} (for a review see
Refs.~\cite{pomeron-book,Dosch}) that the high--energy meson--meson
elastic scattering amplitude can be approximately reconstructed in two
steps. 

i) One first evaluates, in the functional--integral approach, the high--energy
elastic scattering amplitude ${\cal M}_{(dd)}$ of two $q \bar{q}$ pairs
(usually called {\it dipoles}) of given transverse sizes $\vec{R}_{1\perp}$
and $\vec{R}_{2\perp}$ and given longitudinal--momentum fractions $f_1$ and
$f_2$ of the two quarks in the two dipoles respectively.
This {\it dipole--dipole scattering amplitude} turns out to be
governed by the (properly normalised) correlation function of two Wilson loops
${\cal W}_1$ and ${\cal W}_2$, which follow the classical straight lines for
quark (antiquark) trajectories:
\be
{\cal M}_{(dd)} (s,t;\vec{R}_{1\perp},f_1,\vec{R}_{2\perp},f_2) \equiv
-i~2s \displaystyle\int d^2 \vec{z}_\perp
e^{i \vec{q}_\perp \cdot \vec{z}_\perp}
\left[ \dfrac{\langle {\cal W}_1 {\cal W}_2 \rangle}{\langle {\cal W}_1
    \rangle \langle {\cal W}_2 \rangle} -1 \right] , 
\label{scatt-loop}
\ee
where $s$ and $t = -|\vec{q}_\perp|^2$ ($\vec{q}_\perp$ being the transferred
momentum) are the usual Mandelstam variables.
The expectation values $\langle {\cal W}_1 {\cal W}_2 \rangle$,
$\langle {\cal W}_1 \rangle$, $\langle {\cal W}_2 \rangle$ are averages
in the sense of the QCD functional integral ($\langle  {\cal O}[A]
\rangle \equiv \frac{1}{Z} \int [dA] {\cal O}[A]{\rm det}{\cal Q}[A]
e^{iS[A]}$, $Z \equiv \int [dA]{\rm det}{\cal Q}[A] e^{iS[A]}$,  
with $S[A]$ the pure--gauge action and ${\rm det}{\cal Q}[A]$ the
fermion--matrix determinant) and the Wilson loops ${\cal W}_1$ and
${\cal W}_2$ are defined as follows (for $N_c$--colour QCD): 
\be
{\cal W}^{(T)}_{1,2} \equiv
{\dfrac{1}{N_c}} \Tr \left\{ {\cal P} \exp
\left[ -ig \displaystyle\oint_{{\cal C}_{1,2}} A_\mu(x) dx^\mu \right]
\right\} ,
\label{QCDloops}
\ee
where ${\cal P}$ denotes the {\it path ordering} along the given path
${\cal C}$ and $A_\mu = A_\mu^a T^a$, $T^a$ being the generators of
the $SU(N_c)$ Lie algebra in the fundamental representation;
${\cal C}_1$ and ${\cal C}_2$ are two rectangular paths which
follow the classical straight lines for the quark [$X_{q}(\tau)$, forward in
proper time $\tau$] and the antiquark [$X_{\bar{q}}(\tau)$, backward in $\tau$]
trajectories, i.e.,
\bea
{\cal C}_1 &:&
X_{1q}^\mu(\tau) = z^\mu + {p_1^\mu \over m} \tau + (1-f_1) R_1^\mu ,~~~~
X_{1\bar{q}}^\mu(\tau) = z^\mu + {p_1^\mu \over m} \tau - f_1 R_1^\mu ,
\nonumber \\
{\cal C}_2 &:&
X_{2q}^\mu(\tau) = {p_2^\mu \over m} \tau + (1-f_2) R_2^\mu ,~~~~
X_{2\bar{q}}^\mu(\tau) = {p_2^\mu \over m} \tau - f_2 R_2^\mu ,
\label{traj}
\eea
and are closed by straight--line paths at proper times $\tau = \pm T$, where
$T$ plays the role of an infrared (IR) cutoff~\cite{Verlinde,Meggiolaro02},
which can and must be removed in the end, by letting $T \to \infty$. (In fact,
differently from the parton--parton scattering amplitudes, which are known to
be affected by IR divergences, the elastic scattering amplitude of
two colourless states in gauge theories, e.g., two $q \bar{q}$ meson states,
is expected to be an IR--finite physical quantity~\cite{BL}.)\\
Here $p_1$ and $p_2$ are the four--momenta of the two dipoles, taken for
simplicity with the same mass $m$, moving (in the center--of--mass frame)
with speed $V$ and $-V$ along, for example, the $x^1$--direction:
\be
p_1 =
m \left( \cosh {\chi \over 2},\sinh {\chi \over 2},\vec{0}_\perp \right) ,~~~
p_2 =
m \left( \cosh {\chi \over 2},-\sinh {\chi \over 2},\vec{0}_\perp \right) .
\label{p1p2}
\ee
Here $\chi = 2~{\rm arctanh} V$ is the hyperbolic angle between the
two trajectories $1q$ and $2q$, i.e., $p_1 \cdot p_2 = m^2 \cosh\chi$.
In the high--energy limit $s \to \infty$ (i.e., $\chi \to +\infty$),
\be
s \equiv (p_1 + p_2)^2 = 2m^2 \left( \cosh\chi + 1 \right) ~,~~~~{\rm i.e.:}~~
\chi \mathop{\sim}_{s \to \infty} \log \left( {s \over m^2} \right) .
\label{logs}
\ee
Moreover, $R_1 = (0,0,\vec{R}_{1\perp})$, $R_2 = (0,0,\vec{R}_{2\perp})$
and $z = (0,0,\vec{z}_\perp)$, where $\vec{z}_\perp = (z^2,z^3)$ is the
impact--parameter distance between the two loops in the transverse plane.
The two Wilson loops are schematically shown in Fig.~\ref{fig:1}.

If we introduce the following notation for the normalised {\it
  connected} loop--loop correlator (in the presence of a {\it finite} IR cutoff
$T$),~\footnote{The quantity $\mathcal{G}_M$ in Eq.~(\ref{GM}) is equal to
the quantity $\mathcal{G}_M -1$ of
Refs.~\cite{Meggiolaro02,Meggiolaro05,Meggiolaro07,crossing}.}
\be
{\cal G}_M(\chi;T;\vec{z}_\perp;1,2) \equiv
{ \langle {\cal W}^{(T)}_1 {\cal W}^{(T)}_2 \rangle \over
\langle {\cal W}^{(T)}_1 \rangle
\langle {\cal W}^{(T)}_2 \rangle } - 1,
\label{GM}
\ee
where the arguments ``$1$'' and ``$2$'' stand for ``$\vec{R}_{1\perp}, f_1$''
and ``$\vec{R}_{2\perp}, f_2$'' respectively,
the dipole--dipole scattering amplitude (\ref{scatt-loop}) can be rewritten as
\bea
\lefteqn{
 {\cal M}_{(dd)} (s,t;\vec{R}_{1\perp},f_1,\vec{R}_{2\perp},f_2) =}
\nonumber \\
& & -i~2s~\displaystyle\int d^2 \vec{z}_\perp e^{i \vec{q}_\perp \cdot \vec{z}_\perp}
{\cal G}_M (
\chi \mathop{\sim}_{s \to \infty} \log \left( {s \over m^2} \right);
T\to+\infty;\vec{z}_\perp;1,2) .
\label{scatt-loop2}
\eea

ii) The {\it hadron--hadron} (in our case {\it meson--meson}) {\it elastic
scattering amplitude} ${\cal M}_{(hh)}$ can then be obtained by averaging the
dipole--dipole scattering amplitude (\ref{scatt-loop2}) over all possible
dipole transverse separations $\vec{R}_{1\perp}$ and $\vec{R}_{2\perp}$ and
longitudinal--momentum fractions $f_1$ and $f_2$ with two proper squared
hadron wave functions $|\psi_1 (\vec{R}_{1\perp},f_1)|^2$ and
$|\psi_2 (\vec{R}_{2\perp},f_2)|^2$, describing the two interacting mesons:
\bea
{\cal M}_{(hh)}(s,t) &=&
\displaystyle\int d^2\vec{R}_{1\perp} \int_0^1 df_1~
|\psi_1(\vec{R}_{1\perp},f_1)|^2
\displaystyle\int d^2\vec{R}_{2\perp} \int_0^1 df_2~
|\psi_2(\vec{R}_{2\perp},f_2)|^2
\nonumber \\
&\times& {\cal M}_{(dd)} (s,t;\vec{R}_{1\perp},f_1,\vec{R}_{2\perp},f_2) .
\label{scatt-hadron}
\eea
(For the treatment of baryons, a similar, but, of course, more involved,
picture can be adopted, using a genuine three--body configuration or,
alternatively and even more simply, a quark--diquark configuration: we refer
the interested reader to the above--mentioned original
references~\cite{pomeron-book,DFK,Nachtmann97,BN,Dosch}.) 

\begin{figure}[ht]
\centering
\includegraphics[height=0.8\textheight]{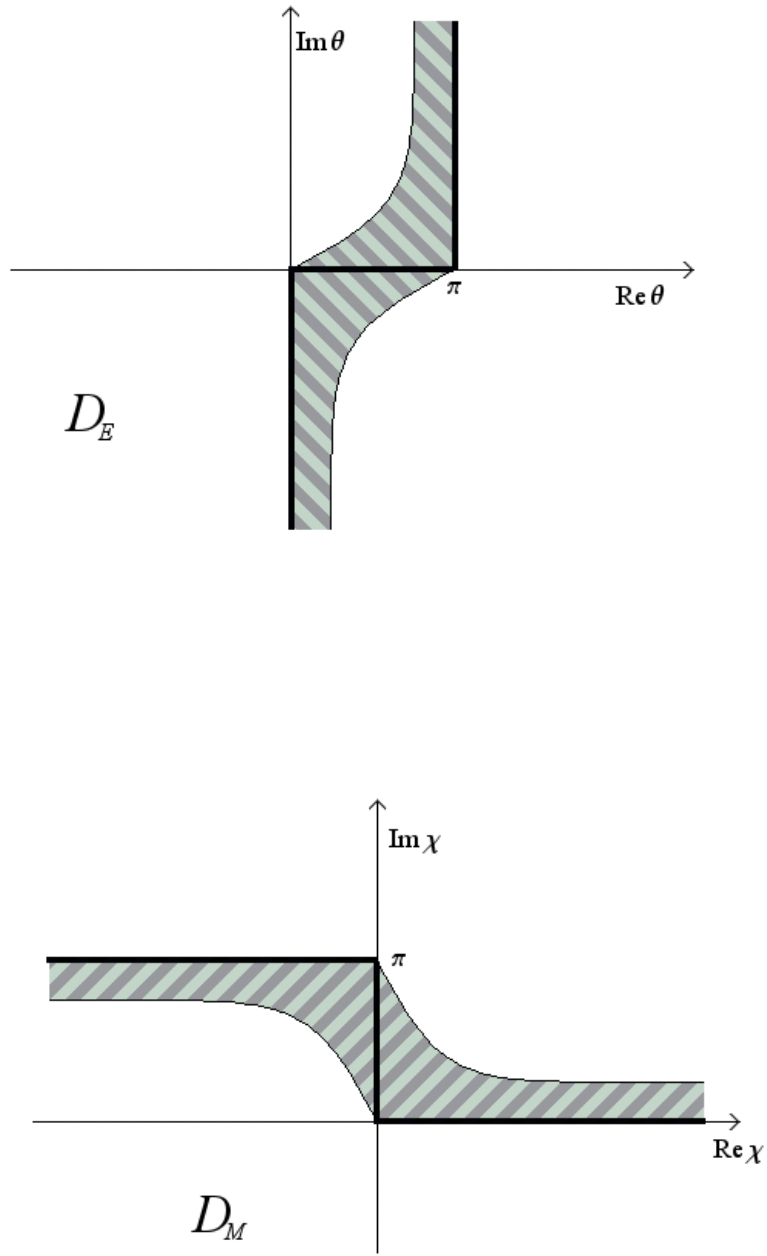}
\vspace{1.5cm}
\caption{The analyticity domains of the function
$\overline{\cal G}_E$ in the complex variable $\theta$ and of the function
$\overline{\cal G}_M$ in the complex variable $\chi$.} 
\label{fig:2}
\end{figure}

All the above refers to the theory in the (physical) Minkowski space--time.
However, as we have already observed in the Introduction,
most of the nonperturbative methods in field theory are available only in the
functional--integral approach formulated in Euclidean space:
therefore we are interested in the problem of the {\it analytic continuation}
of the loop--loop correlation functions into/from Euclidean space.
In the Euclidean theory we can consider the correlation function of two
Euclidean Wilson loops $\widetilde{\cal W}_1$ and $\widetilde{\cal W}_2$,
defined in the same way as they are in Minkowski space--time, taking into
account that the metric is now Euclidean rather than Minkowskian,
and running along two rectangular paths
$\widetilde{\cal C}_1$ and $\widetilde{\cal C}_2$ which follow the following
straight--line trajectories
\bea
\widetilde{\cal C}_1 &:&
X^{1q}_{E\mu}(\tau) = z_{E\mu} + {p_{1E\mu} \over m} \tau
+ (1-f_1) R_{1E\mu} ,~~~~
X^{1\bar{q}}_{E\mu}(\tau) = z_{E\mu} + {p_{1E\mu} \over m} \tau 
- f_1 R_{1E\mu} ,
\nonumber \\
\widetilde{\cal C}_2 &:&
X^{2q}_{E\mu}(\tau) = {p_{2E\mu} \over m} \tau + (1-f_2) R_{2E\mu} ,~~~~
X^{2\bar{q}}_{E\mu}(\tau) = {p_{2E\mu} \over m} \tau - f_2 R_{2E\mu} ,
\label{trajE}
\eea
and are closed by straight--line paths at proper times $\tau = \pm
T$. Here the Euclidean coordinates are
$X_{E\mu}=(X_{E1},X_{E2},X_{E3},X_{E4})$, where $X_{E4}$ corresponds to
the ``Euclidean time'', and $R_{1E} = (0,\vec{R}_{1\perp},0)$, $R_{2E} =
(0,\vec{R}_{2\perp},0)$ and  $z_E = (0,\vec{z}_\perp,0)$, where
$\vec{R}_{1\perp}$, $\vec{R}_{2\perp}$ and $\vec{z}_{\perp}$ are
exactly the same transverse vectors introduced in the Minkowskian
case. Moreover, in the Euclidean theory we {\it choose} the
four--vectors $p_{1E}$ and $p_{2E}$ to be
\be
p_{1E} =
m \left( \sin{\theta \over 2}, \vec{0}_\perp, \cos{\theta \over 2} \right) ,~~~
p_{2E} =
m \left( -\sin{\theta \over 2}, \vec{0}_\perp, \cos{\theta \over 2} \right) ,
\label{p1p2E}
\ee
$\theta$ being the angle formed by the two trajectories $1q$ and $2q$ in
Euclidean four--space, i.e., $p_{1E} \cdot p_{2E} = m^2 \cos\theta$.\\
It turns out~\cite{Meggiolaro02,Meggiolaro05} that the Minkowskian
quantity ${\cal G}_M$ [with $\chi\in\mathbb{R}^+$] can be 
reconstructed from the corresponding Euclidean quantity [with $\theta \in (0,\pi)$] 
\be
{\cal G}_E(\theta;T;\vec{z}_\perp;1,2) \equiv
{ \langle \widetilde{\cal W}^{(T)}_1 \widetilde{\cal W}^{(T)}_2 \rangle \over
\langle \widetilde{\cal W}^{(T)}_1 \rangle
\langle \widetilde{\cal W}^{(T)}_2 \rangle } - 1,
\label{GE}
\ee
(where $\langle \ldots \rangle$ are now averages
in the sense of the Euclidean functional integral, i.e., \linebreak
$\langle  {\cal O}[A_E]
\rangle \equiv \frac{1}{Z_E} \int [dA_E] {\cal O}[A_E]{\rm det}{\cal Q}_E[A_E]
e^{-S_E[A_E]}$, $Z_E \equiv \int [dA_E]{\rm det}{\cal Q}_E[A_E]
e^{-S_E[A_E]}$, with $S_E[A_E]$ the pure--gauge Euclidean action and ${\rm
  det}{\cal Q}_E[A_E]$ the Euclidean fermion--matrix determinant)\footnote{We 
  note here that, exploiting the invariance 
  of the theory under charge conjugation ($C$), one can show that the
  correlation function ${\cal G}_E$, as well as the quantity ${\cal
    C}_E$ to be defined below in Eq.~(\ref{C12}), is a {\it real}
  quantity (differently from the corresponding Minkowskian
  correlator). In fact, under a $C$ transformation, $A_{E\mu}\to
  A_{E\mu}^{(c)} = -A_{E\mu}^{t} = -A_{E\mu}^*$, the integration
  measure $[dA_E]$, the gauge--field action $S_E$ and the fermion
  matrix determinant ${\rm det}{\cal Q}_E[A_E]$ (which are {\it real}
  quantities) are invariant, while a Wilson loop $\widetilde{\cal W}$
  goes into its complex conjugate $\widetilde{\cal W}^*$.} 
by an analytic continuation in the angular
variables $\theta \to -i\chi$ and in the IR cutoff $T \to iT$, exactly as
in the case of Wilson lines~\cite{Meggiolaro97,Meggiolaro98,Meggiolaro02}.
This result~\cite{Meggiolaro02,Meggiolaro05} is derived under certain
hypotheses of analyticity in the angular variables and in the IR cutoff $T$.
In particular, one makes the assumption~\cite{crossing} that
the function ${\cal G}_E$, as a function of the {\it complex}
variable $\theta$, can be {\it analytically extended} from the real segment
$(0 < \re\theta < \pi, \im\theta = 0)$ to a domain ${\cal D}_E$,
which also includes the negative imaginary axis
$(\re\theta = 0+, \im\theta < 0)$; therefore,
the function ${\cal G}_M$, as a function of the {\it complex} variable
$\chi$, can be {\it analytically extended} from the positive
real axis $(\re\chi > 0, \im\chi = 0+)$ to a domain
${\cal D}_M = \{ \chi \in \mathbb{C} ~|~ -i\chi \in {\cal D}_E \}$,
which also includes the imaginary segment $(\re\chi = 0, 0 < \im\chi < \pi)$.
The validity of this assumption is confirmed by explicit calculations in
perturbation theory~\cite{Meggiolaro97,Meggiolaro05,BB}.
The domains ${\cal D}_E$ and ${\cal D}_M$ are schematically shown in
Fig.~\ref{fig:2}.
Denoting with $\overline{\cal G}_M$ and $\overline{\cal G}_E$ such analytic
extensions, we then have the following {\it analytic--continuation
  relations}~\cite{Meggiolaro05,crossing}:
\bea
\overline{\cal G}_E(\theta;T;\vec{z}_\perp;1,2)
&=& \overline{\cal G}_M (i\theta;-iT;\vec{z}_\perp;1,2) ,
\qquad \forall\theta\in {\cal D}_E ;
\nonumber \\
\overline{\cal G}_M(\chi;T;\vec{z}_\perp;1,2)
&=& \overline{\cal G}_E (-i\chi;iT;\vec{z}_\perp;1,2) ,
\qquad \forall\chi\in {\cal D}_M .
\label{analytic}
\eea
As we have said above, the loop--loop correlation functions
(both in the Minkowskian and in the Euclidean theories) are expected to be
IR--{\it finite} quantities, i.e., to have finite limits when $T \to \infty$,
differently from what happens in the case of Wilson lines.
One can then define the following loop--loop correlation functions
with the IR cutoff removed:
\bea
{\cal C}_M(\chi;\vec{z}_\perp;1,2) &\equiv&
\displaystyle\lim_{T \to \infty}
{\cal G}_M(\chi;T;\vec{z}_\perp;1,2) , \nonumber \\
{\cal C}_E(\theta;\vec{z}_\perp;1,2)
&\equiv& \displaystyle\lim_{T \to \infty}
{\cal G}_E(\theta;T;\vec{z}_\perp;1,2) .
\label{C12}
\eea
It has been proved in Ref.~\cite{Meggiolaro05} that, under certain analyticity
conditions in the {\it complex} variable $T$ [conditions which are also
sufficient to make the relations (\ref{analytic}) meaningful], the two
quantities (\ref{C12}), obtained {\it after} the removal of the IR cutoff
($T \to \infty$), are still connected by the usual analytic continuation in
the angular variables only:
\bea
\overline{\cal C}_E(\theta;\vec{z}_\perp;1,2) &=&
\overline{\cal C}_M(i\theta;\vec{z}_\perp;1,2) ,
\qquad \phantom{-}\forall\theta\in {\cal D}_E ;
\nonumber \\
\overline{\cal C}_M(\chi;\vec{z}_\perp;1,2) &=&
\overline{\cal C}_E(-i\chi;\vec{z}_\perp;1,2) ,
\qquad \forall\chi\in {\cal D}_M .
\label{final}
\eea
This is a highly non--trivial result, whose general validity is discussed
in Ref.~\cite{Meggiolaro05}.
The validity of the relation (\ref{final}) for the loop--loop correlators 
in QCD has been also verified in Ref.~\cite{BB} by an explicit
calculation up to the order ${\cal O}(g^6)$ in perturbation theory.
However we want to stress that the analytic continuation (\ref{analytic}) or
(\ref{final}) is expected to be an {\it exact} result, i.e., not restricted
to some order in perturbation theory or to some other approximation,
and is valid both for the Abelian and the non--Abelian cases.

It has also been recently shown in Refs.~\cite{crossing,Meggiolaro07} that the
analytic--continuation relations (\ref{analytic})
allow us to deduce non--trivial properties of the Euclidean correlator
$\mathcal{G}_E$ under the exchange $\theta\to \pi -\theta$ and of the
Minkowskian correlator $\mathcal{G}_M$ under the exchange $\chi\to
i\pi -\chi$, corresponding to the exchange from a loop--loop
correlator to a loop--{\it antiloop} correlator, where an {\it
  antiloop} is obtained from a given loop by exchanging the quark and
the antiquark trajectories:
\bea
\label{crossing}
&\mathcal{G}_E(\pi-\theta;T;\vec{z}_{\perp};1,2)
=\mathcal{G}_E(\theta;T;\vec{z}_{\perp};1,\overline{2}) 
=\mathcal{G}_E(\theta;T;\vec{z}_{\perp};\overline{1},2) ,
&\forall\theta\in (0,\pi) ;
\nonumber \\
&\overline{\mathcal{G}}_M(i\pi-\chi;T;\vec{z}_{\perp};1,2)
=\mathcal{G}_M(\chi;T;\vec{z}_{\perp};1,\overline{2}) 
=\mathcal{G}_M(\chi;T;\vec{z}_{\perp};\overline{1},2) ,
&\forall\chi\in\mathbb{R}^+,
\eea
where the arguments ``$\overline{1}$'' and ``$\overline{2}$'' stand
for ``$-\vec{R}_{1\perp}, 1-f_1$'' 
and ``$-\vec{R}_{2\perp}, 1-f_2$'' respectively.
These two relations are known as {\it crossing--symmetry relations}
for loop--loop correlators. As they are valid for every value of the
IR cutoff $T$, completely analogous relations also hold for the
loop--loop correlation functions ${\cal C}_M$ and ${\cal C}_E$ with
the IR cutoff removed ($T \to \infty$), defined in Eq.~(\ref{C12}):
\bea
\label{crossingC}
&\mathcal{C}_E(\pi-\theta;\vec{z}_{\perp};1,2)
=\mathcal{C}_E(\theta;\vec{z}_{\perp};1,\overline{2}) 
=\mathcal{C}_E(\theta;\vec{z}_{\perp};\overline{1},2) ,
&\forall\theta\in (0,\pi) ;
\nonumber \\
&\overline{\mathcal{C}}_M(i\pi-\chi;\vec{z}_{\perp};1,2)
=\mathcal{C}_M(\chi;\vec{z}_{\perp};1,\overline{2}) 
=\mathcal{C}_M(\chi;\vec{z}_{\perp};\overline{1},2) ,
&\forall\chi\in\mathbb{R}^+.
\eea

Taking into account the aforementioned analytic--continuation relations,
we can rewrite Eq.~(\ref{scatt-hadron}) in terms of the Euclidean
correlation function ${\cal C}_E$ as
\bea
  \label{eq:Eamp}
  {\cal M}_{(hh)}(s,t) &=&
  -i~2s\displaystyle\int d^2\vec{R}_{1\perp} \int_0^1 df_1~
  |\psi_1(\vec{R}_{1\perp},f_1)|^2
  \displaystyle\int d^2\vec{R}_{2\perp} \int_0^1 df_2~
  |\psi_2(\vec{R}_{2\perp},f_2)|^2
  \nonumber \\
  & &\times \displaystyle\int d^2\vec{z}_{\perp}
  e^{i\vec{q}_{\perp}\cdot\vec{z}_{\perp}}  {\cal C}_{E} (\theta\to
  -i\log\left(\dfrac{s}{m^2}\right);\vec{z}_{\perp};\vec{R}_{1\perp},f_1,\vec{R}_{2\perp},f_2).
\eea
By virtue of the optical theorem, Eq.~(\ref{optical}), the total cross
section is then given by the expression
\bea
  \label{eq:WLoptic0}
  \sigma_{\rm tot}^{(hh)} (s)  \mathop{\sim}_{s \to \infty} &
  -2\displaystyle\int d^2\vec{R}_{1\perp} \int_0^1 df_1~
  |\psi_1(\vec{R}_{1\perp},f_1)|^2
  \displaystyle\int d^2\vec{R}_{2\perp} \int_0^1 df_2~
  |\psi_2(\vec{R}_{2\perp},f_2)|^2
  \nonumber \\
&  \times \int d^2z_{\perp} {\,\rm Re\,}{\cal C}_{E} (\theta\to
  -i\log\left(\dfrac{s}{m^2}\right);\vec{z}_{\perp};\vec{R}_{1\perp},f_1,\vec{R}_{2\perp},f_2). 
\eea
If one chooses hadron wave functions invariant under rotations and
under the exchange $f_i\to 1-f_i$ (see Refs.~\cite{Dosch,LLCM1} and
also~\cite{pomeron-book}, \S 8.6, and references therein), the
correlation function ${\cal C}_E$ in Eqs.~(\ref{eq:Eamp}) and
(\ref{eq:WLoptic0}) can be substituted (without changing the result)
with the following {\it averaged} correlation function:
\bea
\lefteqn{ {\cal
    C}_E^{ave}(\theta;\vec{z}_{\perp};|\vec{R}_{1\perp}|,f_1,|\vec{R}_{2\perp}|,f_2)
  \equiv
    \int d\hat{R}_{1\perp} \int d\hat{R}_{2\perp}  }
    \nonumber 
    \label{eq:ave} \\
    & &\times\frac{1}{4}\left\{
      {\cal
        C}_E(\theta;\vec{z}_{\perp};\vec{R}_{1\perp},f_1,\vec{R}_{2\perp},f_2)+ 
      {\cal
        C}_E(\theta;\vec{z}_{\perp};\vec{R}_{1\perp},1-f_1,\vec{R}_{2\perp},f_2)\right. \nonumber \\ 
    & &+ \left. {\cal
        C}_E(\theta;\vec{z}_{\perp};\vec{R}_{1\perp},f_1,\vec{R}_{2\perp},1-f_2) 
      + {\cal
        C}_E(\theta;\vec{z}_{\perp};\vec{R}_{1\perp},1-f_1,\vec{R}_{2\perp},1-f_2)\right\},
\eea
where $\int d\hat{R}_{i\perp}$ stands for integration over the
orientations of $\vec{R}_{i\perp}$. We note here that,
as a consequence of the {\it crossing--symmetry relations}
Eq.~(\ref{crossingC}), the function ${\cal C}_E^{ave}$  
is automatically {\it crossing--symmetric}, i.e., ${\cal C}_E^{ave}(\pi-\theta;\ldots)={\cal
  C}_E^{ave}(\theta;\ldots)$ for fixed values of the other variables.

In the following, if not specified otherwise, we will take for simplicity
the longitudinal--momentum fractions $f_1$ and $f_2$ of the two quarks in the
two dipoles (and, therefore, also the longitudinal--momentum fractions $1-f_1$
and $1-f_2$ of the two antiquarks in the two dipoles) to be fixed to $1/2$:
this is known to be a good approximation for hadron--hadron interactions (see
Refs.~\cite{pomeron-book,Dosch} and references therein). We will also adopt
the notation ${\cal
  G}_E(\theta;T;\vec{z}_{\perp};\vec{R}_{1\perp},\vec{R}_{2\perp})\equiv
{\cal G}_E(\theta;T;\vec{z}_{\perp};\vec{R}_{1\perp},f_1={1\over 
  2},\vec{R}_{2\perp},f_2={1\over 2})$, 
and similarly for ${\cal C}_E$ and ${\cal C}_E^{ave}$. Note that in
this case the quantity ${\cal C}_E^{ave}$ reduces to the average over
the transverse orientations only:
\be
  \label{eq:ave2}
  {\cal C}_E^{ave}(\theta;\vec{z}_{\perp};|\vec{R}_{1\perp}|,|\vec{R}_{2\perp}|)=
    \int d\hat{R}_{1\perp} \int d\hat{R}_{2\perp}~{\cal
      C}_E(\theta;\vec{z}_{\perp};\vec{R}_{1\perp},\vec{R}_{2\perp}). 
\ee
In the next section we will show how it is possible to calculate these
Euclidean correlation functions in LGT and discuss the main technical
difficulties of this approach.

\newsection{Loop--loop correlators on the lattice}

Being a gauge--invariant quantity, the Wilson--loop correlation function
${\cal G}_E$ is a natural candidate for a lattice computation;
nevertheless, a number of complications arise because of the explicit
breaking of $O(4)$ invariance on a lattice. The major complication is
due to the limited number of possible orientations of rectangular
loops on the lattice. As straight lines on a hypercubic lattice can be
either parallel or orthogonal, the values of $\theta$ directly
accessible are limited to $0^{\circ}$, $90^{\circ}$ and
$180^{\circ}$. To cover a significantly large set of angles, we then
have to make use of {\it off--axis} (and so non planar) Wilson loops,
thus introducing in our approach another approximation  that should be
carefully discussed. 

\begin{figure}
  \begin{center}
    {\includegraphics{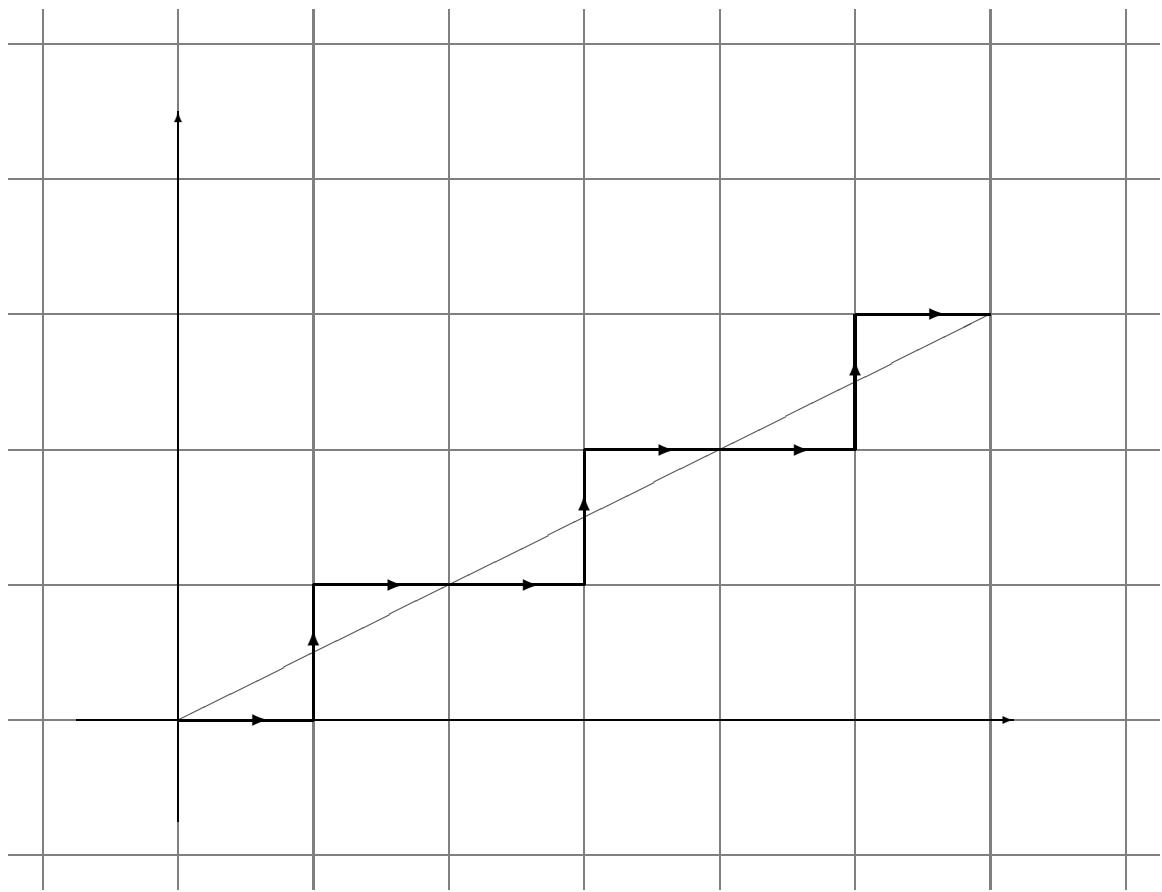}}
    \vspace{1.5cm}
    \caption{The minimal--distance prescription for a line with slope $\tan\theta
      = 1/2$.}
    \label{fig:Brespres}
  \end{center}
\end{figure}

The loops involved in the calculation of ${\cal G}_E$ have one side
(from now on the {\it longitudinal} side) in the $(x_{E1},x_{E4})$
(longitudinal) plane and the other ({\it transverse} side) in the
$(x_{E2},x_{E3})$ (transverse) plane and their centers are separated in the
transverse plane, so that the problem of reproducing the loop
configuration consists effectively of two distinct two--dimensional
problems. It is reasonable to evaluate the loop sides on the lattice paths
that minimise the distance from the true, continuum paths, in order to stay as
``close'' as possible to the continuum limit, adopting essentially the same
strategy as in computer graphics when drawing straight lines on a
screen (see Fig.~\ref{fig:Brespres}); such  ``{\it minimal--distance paths}''
can be found in a very efficient way by means of the well--known {\it
  Bresenham algorithm}~\cite{Bres}, which has already been used in lattice 
calculations (see e.g.~\cite{Bali}, where it is also generalised to
the three--dimensional case). To every continuum straight
path having as endpoints two coplanar lattice points we can then
associate unambiguously a Wilson line on the lattice by means of the
minimal--distance prescription\footnote{For lines inclined at
  $45^{\circ}$ with respect to an axis a certain ambiguity remains,
  but we can average over equivalent paths.} and then build with them
the Wilson loops we are interested in. Such loops are identified by
the position of the center and by two two--dimensional lattice vectors;
we thus define  $\widetilde{\cal
  W}_L(\vec{l}_{\parallel};\vec{r}_{\perp};n)$ to be the lattice
Wilson loop evaluated on the minimal path that approximates the 
rectangle having as corners the lattice points $n-l/2-r/2$,
$n+l/2-r/2$, $n+l/2+r/2$ and  $n-l/2+r/2$, where $n$, $l$ and $r$ are
vectors in lattice units, $n$ is the 
position of the center and\footnote{The components of $l$ and $r$ must
  be integers. Note that the center of the loop may not lie on a
  lattice point.} $\vec{l}_{\parallel} =
(l_1,l_4)$, $l_2=l_3=0$, $\vec{r}_{\perp}  = (r_2,r_3)$, $r_1=r_4=0$.

On the lattice we then define the correlator 
\be
\label{eq:corr_lat}
{\cal
  G}_L(\vec{l}_{1\parallel},\vec{l}_{2\parallel};\vec{d}_{\perp};\vec{r}_{1\perp},\vec{r}_{2\perp})
\equiv \frac{\langle\widetilde{\cal 
    W}_L(\vec{l}_{1\parallel};\vec{r}_{1\perp};d)\widetilde{\cal
    W}_L(\vec{l}_{2\parallel};\vec{r}_{2\perp};0)\rangle}{\langle\widetilde{\cal 
  W}_L(\vec{l}_{1\parallel};\vec{r}_{1\perp};d)\rangle\langle\widetilde{\cal
  W}_L(\vec{l}_{2\parallel};\vec{r}_{2\perp};0)\rangle} - 1,
\ee
where $d=(0,\vec{d}_{\perp},0)$, $\vec{d}_{\perp} = (d_2,d_3)$. As (full) $O(4)$
invariance is broken, this correlator depends explicitly on the
(two--dimensional) lattice vectors $\vec{l}_{i\parallel}$,
$\vec{r}_{i\perp}$ ($i=1,2$) and $\vec{d}_{\perp}$ rather than on
their scalar products: indeed, for a given relative orientation we can
find different realisations in terms of lattice vectors, generally
inequivalent as they involve different Wilson--loop operators. Anyway,
as rotation invariance is restored in the continuum limit, we expect 
\be
\label{eq:contlim}
{\cal
  G}_L(\vec{l}_{1\parallel},\vec{l}_{2\parallel};\vec{d}_{\perp};\vec{r}_{1\perp},\vec{r}_{2\perp}) 
  \mathop\simeq_{a\to 0} {\cal
  G}_E(\theta;T_1=aL_1/2,T_2=aL_2/2;a\vec{d}_{\perp};a\vec{r}_{1\perp},a\vec{r}_{2\perp}),
\ee
where $L_i \equiv |\vec{l}_{i\parallel}|$ are what we define to be the
lengths of the longitudinal sides of the loops in lattice units (from
now on, ``{\it lengths}''), and
$\vec{l}_{1\parallel}\cdot\vec{l}_{2\parallel} \equiv L_1 L_2
  \cos\theta$. As it is impossible, in general, to have 
\mbox{$L_1=L_2$}, we  have relaxed this condition and have considered
in Eq.~(\ref{eq:contlim}) the correlation function with two
IR cutoffs $T_1$ and $T_2$,
\be
{\cal G}_E(\theta;T_1,T_2;\vec{z}_{\perp};\vec{R}_{1\perp},\vec{R}_{2\perp}) \equiv
\frac{\langle \widetilde{{\cal W}}^{(T_1)}_1 \widetilde{{\cal
      W}}^{(T_2)}_2\rangle}{\langle \widetilde{{\cal W}}^{(T_1)}_1\rangle
  \langle\widetilde{{\cal W}}^{(T_2)}_2\rangle} - 1,
\ee
where with a small abuse of notation we have kept the same notation for the
correlation function as in Eq.~(\ref{GE}). As already pointed out in
Section 2, the correlator ${\cal G}_E$ is expected to be an IR--finite
quantity, so that we can regularise it considering loops with different
lengths $T_1$ and $T_2$ and taking the limits $T_1,T_2\to\infty$
independently, obtaining the same function ${\cal C}_E$ defined in Eq.~(\ref{C12}).

On top of these field theoretical complications we have to face the numerical
difficulty (or even feasibility) of a lattice ``measurement'' of the
relevant correlation function of Wilson loops. The Wilson--loop expectation
value is known to obey an ``area law'', i.e., to vanish exponentially with
its area for large areas; moreover, in the 't Hooft large-$N_c$ expansion with
$g^2N_c$ kept constant such correlators are known to factorise to
leading order, thus giving
\be
{\cal G}_E = \frac{\langle\widetilde{\cal W}_1^{(T)} \widetilde{\cal
    W}_2^{(T)}\rangle}{\langle\widetilde{\cal W}_1^{(T)}\rangle \langle\widetilde{\cal
    W}_2^{(T)}\rangle} - 1 = \frac{\langle\widetilde{\cal
    W}_1^{(T)}\rangle\langle\widetilde{\cal W}_2^{(T)}\rangle + {\cal
    O}(\frac{1}{N_c^2})}{\langle\widetilde{\cal W}_1^{(T)}\rangle
    \langle\widetilde{\cal W}_2^{(T)}\rangle} - 1 = {\cal O}(\frac{1}{N_c^2}).
\ee
We then expect ${\cal G}_E$ to be a small quantity, obtained from the ratio
of two exponentially decreasing quantities as $T$ becomes large. Moreover,
general considerations make plausible a behaviour of the kind ${\cal
  G}_E \sim e^{-\gamma|\vec{z}_{\perp}|}$ at
large separations, and thus we expect the noise to overcome the signal
after a few lattice spacings.  

We can try to maximise the information obtained from each thermalised
configuration, in order to reduce the statistical noise, by exploiting
the symmetries of the lattice. 
It is easy to see that the chosen prescription for the construction of
loops is consistent with lattice rotations and reflections [i.e., the
  cubic subgroup of $O(4)$], in the following sense. One can perform
a cubic transformation on a Wilson loop (in the {\it continuum}), and
then construct its {\it lattice} approximation with the given
prescription, or alternatively construct first the {\it lattice}
approximation and then perform the same cubic transformation (this
time on the {\it lattice}), and in both cases one would obtain the
same result. We can then average over cubic transformations of the
whole loop configuration and, imposing periodic boundary conditions,
we can average over lattice translations as well; to further clarify
the numerical signal we can also use the identities 
\bea
\label{eq:lat_id}
& &{\cal G}_L( 
\vec{l}_{1\parallel},\vec{l}_{2\parallel};\vec{d}_{\perp};\vec{r}_{1\perp},\vec{r}_{2\perp})
=\,{\cal
  G}_L(\vec{l}_{1\parallel},\vec{l}_{2\parallel};-\vec{d}_{\perp};-\vec{r}_{1\perp},-\vec{r}_{2\perp})
\nonumber \\ 
& &={\cal
  G}_L(\vec{l}_{1\parallel},\vec{l}_{2\parallel};-\vec{d}_{\perp};\vec{r}_{1\perp},\vec{r}_{2\perp})=
{\cal
  G}_L(\vec{l}_{1\parallel},\vec{l}_{2\parallel};\vec{d}_{\perp};-\vec{r}_{1\perp},-\vec{r}_{2\perp}),
\eea
which can be proved using invariance under cubic symmetry and the
trivial fact that 
\be
\widetilde{\cal W}_L(\vec{l}_{\parallel};-\vec{r}_{\perp};n) = \widetilde{\cal
  W}_L(-\vec{l}_{\parallel};\vec{r}_{\perp};n).  
\ee

\newsection{Numerical results}

We have performed a Monte Carlo calculation of the correlation function
${\cal G}_L$ of two Wilson loops for several values of the relative
angle, various lengths and different configurations in the
transverse plane, on a $16^4$ hypercubic lattice with periodic
boundary conditions. The link
configurations were generated by means of a mixture of
(pseudo)--heat\-bath~\cite{Creutz1,Cab_Mar,KP} and
overrelaxation steps~\cite{Creutz2} with the usual Wilson action for
$SU(3)$ pure--gauge theory~\cite{Wilson}, also
known in the literature as the {\it quenched} approximation of QCD, which
consists in neglecting dynamical fermion loops by setting the fermion
matrix determinant to a constant. Even though, of course, we cannot
exclude that the inclusion of dynamical--fermion effects, via the
fermion--matrix determinant, could introduce new features in the data,
in this paper (which, we want to stress, is the first to approach this
problem from the point of view of lattice QCD) we have preferred to
make the easiest and most convenient choice, that gives us the possibility
of collecting large statistics (also considering the various
difficulties in measuring the correlation functions ${\cal G}_L$, as
explained in the previous section).

We have measured the correlation functions $\langle \widetilde{\cal
  W}_{L1} \widetilde{\cal W}_{L2} \rangle$ and the loop expectation
values $\langle \widetilde{\cal W}_{L1}\rangle$ and  $\langle
\widetilde{\cal  W}_{L2} \rangle$, with \mbox{$\widetilde{\cal
    W}_{L1}\equiv\widetilde{\cal
    W}_{L}(\vec{l}_{1\parallel};\vec{r}_{1\perp};d)$} and
\mbox{$\widetilde{\cal W}_{L2}\equiv\widetilde{\cal
    W}_{L}(\vec{l}_{2\parallel};\vec{r}_{2\perp};0)$}, on 30000 
thermalised configurations at $\beta\equiv 6/g^2=6.0$. 
As it is well known, the lattice spacing $a$ is related to the bare coupling
constant $g$ (i.e., to $\beta$) through the renormalisation group
equation. The lattice scale, i.e., the value of $a$ in physical units, is
determined from the physical value of some relevant (dimensionful) observable
like the string tension or the static $q\bar{q}$ force at some fixed distance
(``{\it Sommer scale}'') (see e.g. Ref.~\cite{Mic_Tep} and 
references therein): in our case  one finds that $a(\beta=6.0)\simeq 0.1\,{\rm fm}$.
The choice of $\beta=6.0$ on a $16^4$ lattice is made in
order to stay within the so--called ``{\it scaling window}'': in this sense we
are relying in an indirect way on the validity of the relation
(\ref{eq:contlim}) between Wilson--loop correlation functions on 
the lattice and in the continuum (and therefore we shall use the notation
${\cal G}_E$/${\cal C}_E$ of the continuum in all the figures reporting our
lattice data). An explicit test of scaling in our
case is more difficult, as one has to keep a large number of length scales
under control while varying the lattice spacing. A possibility could be to
integrate over the distance and the sizes of the loops and then study the
scaling properties of the resulting quantity, although this seems to be a very
hard task. 

\begin{figure}
  \begin{center}
    {\includegraphics{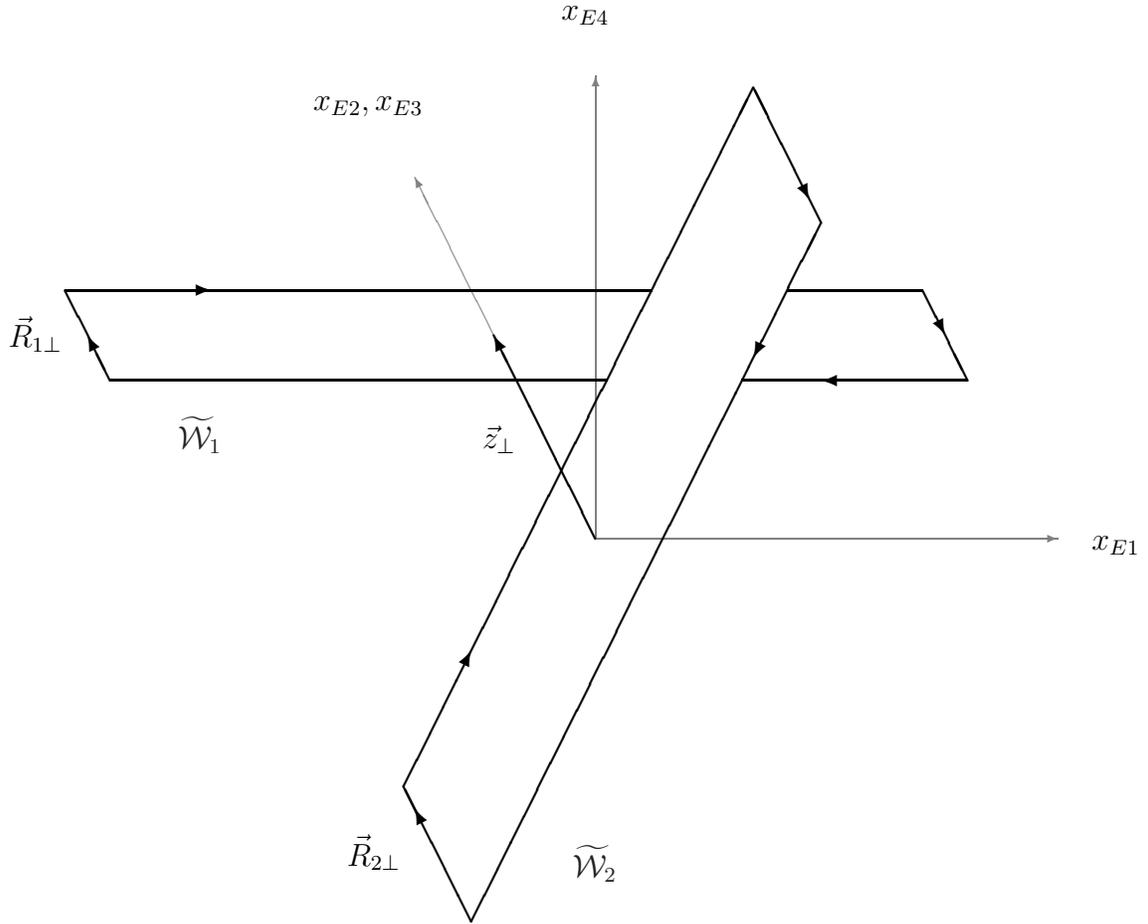}}
    \vspace{1.5cm}
    \caption{The relevant Wilson--loop configuration. Using the $O(4)$
      invariance of the Euclidean theory we have put $p_{1E}$ parallel to
      the $x_{E1}$ axis.}
    \label{fig:loopconf}
  \end{center}
\end{figure}

\begin{figure}
  \begin{center}
    {\includegraphics{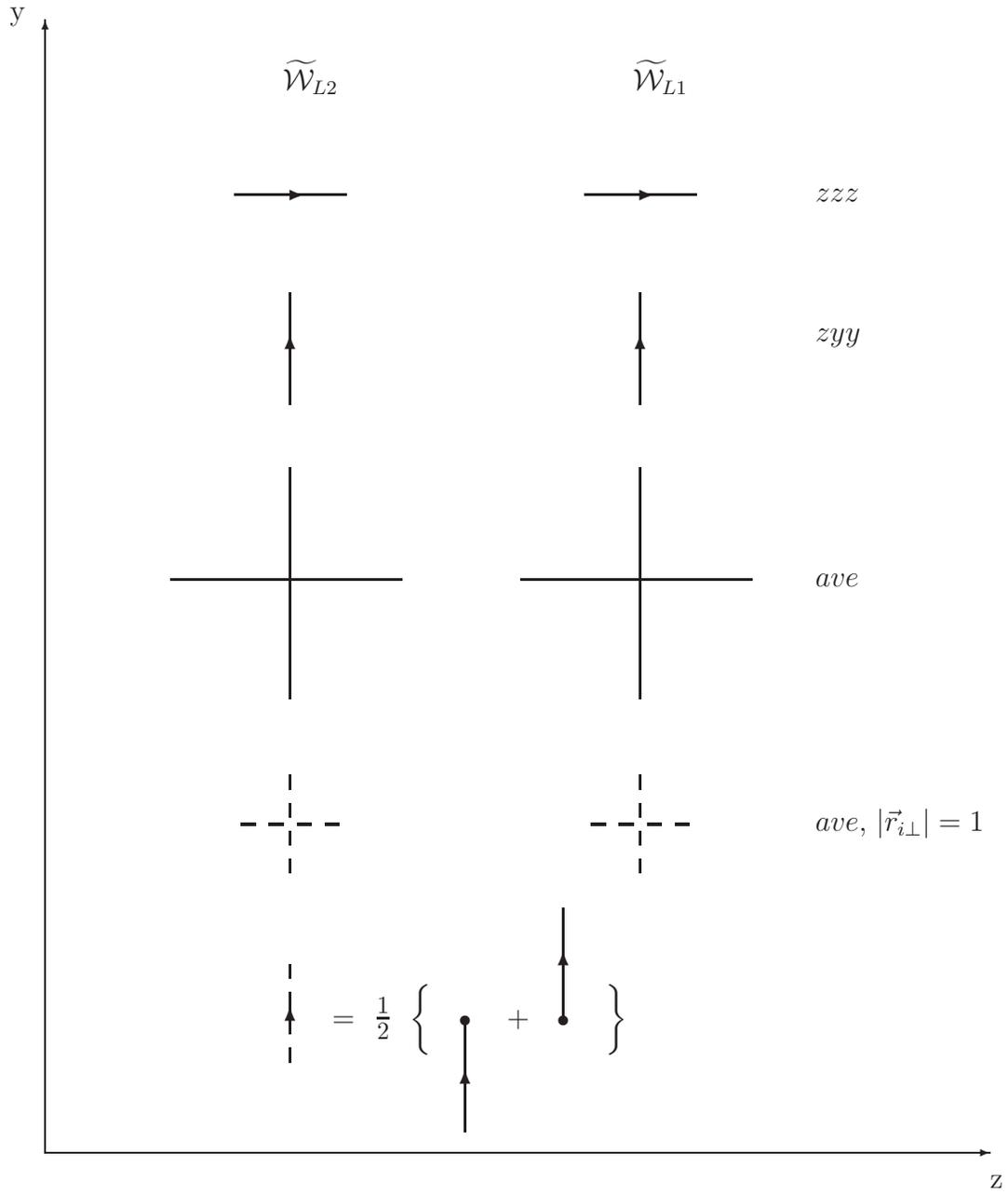}}
    \vspace{1.5cm}
    \caption{Loop configuration in the transverse plane. In the
      ``{\it ave}'' case the link orientation is not shown as it is averaged
    over.}
    \label{fig:transvconf}
  \end{center}
\end{figure}

To keep the corrections due to $O(4)$ invariance breaking as small as
possible, we have kept one of the two loops {\it on--axis} and 
we have only tilted the other one as shown in Fig.~\ref{fig:loopconf};
the on--axis loop  $\widetilde{\cal W}_{L1}$ is taken to be parallel
to the $x_{E1}$ axis, $\vec{l}_{1\parallel}=(L_1,0)$, and of length
$L_1=6,8$. We have used two sets of off--axis loops $\widetilde{\cal
  W}_{L2}$ tilted at $\tan^{-1}(1/2)\simeq 26.565^{\circ}$ and
$45^{\circ}$ with respect to one of the longitudinal axes; the
corresponding lattice vectors $\vec{l}_{2\parallel}$ are listed in
Table \ref{tab:L2}, together with their length and the angle formed
with $\vec{l}_{1\parallel}$. We have used loops with  transverse size
$|\vec{r}_{1\perp}|=|\vec{r}_{2\perp}|=1$ in lattice units; the loop
configurations in the transverse plane are those illustrated in
Fig.~\ref{fig:transvconf}, namely 
$\vec{d}_{\perp} \parallel \vec{r}_{1\perp} \parallel
\vec{r}_{2\perp}$ (which we call ``{\it zzz}'') and $\vec{d}_{\perp} \perp
\vec{r}_{1\perp} \parallel \vec{r}_{2\perp}$  (``{\it zyy}''). As
explained in Section 2, it is interesting to also measure the
orientation--averaged quantity (``{\it ave}'') defined in
Eq.~(\ref{eq:ave2}). The lattice version of this equation is easily
recovered for even (integer) values of the transverse sizes; in our
particular case, $|\vec{r}_{i\perp}|=1$, we have to use a sort of
``{\it smearing}'' procedure, averaging nearby loops as depicted in
Fig.~\ref{fig:transvconf}. Note that in doing so we are actually
averaging over the orientations and over the values $f_i=0,1$
($i=1,2$) of the longitudinal--momentum fractions, according to
Eq.~(\ref{eq:ave}).

\begin{figure}
  \begin{center}
    {\includegraphics{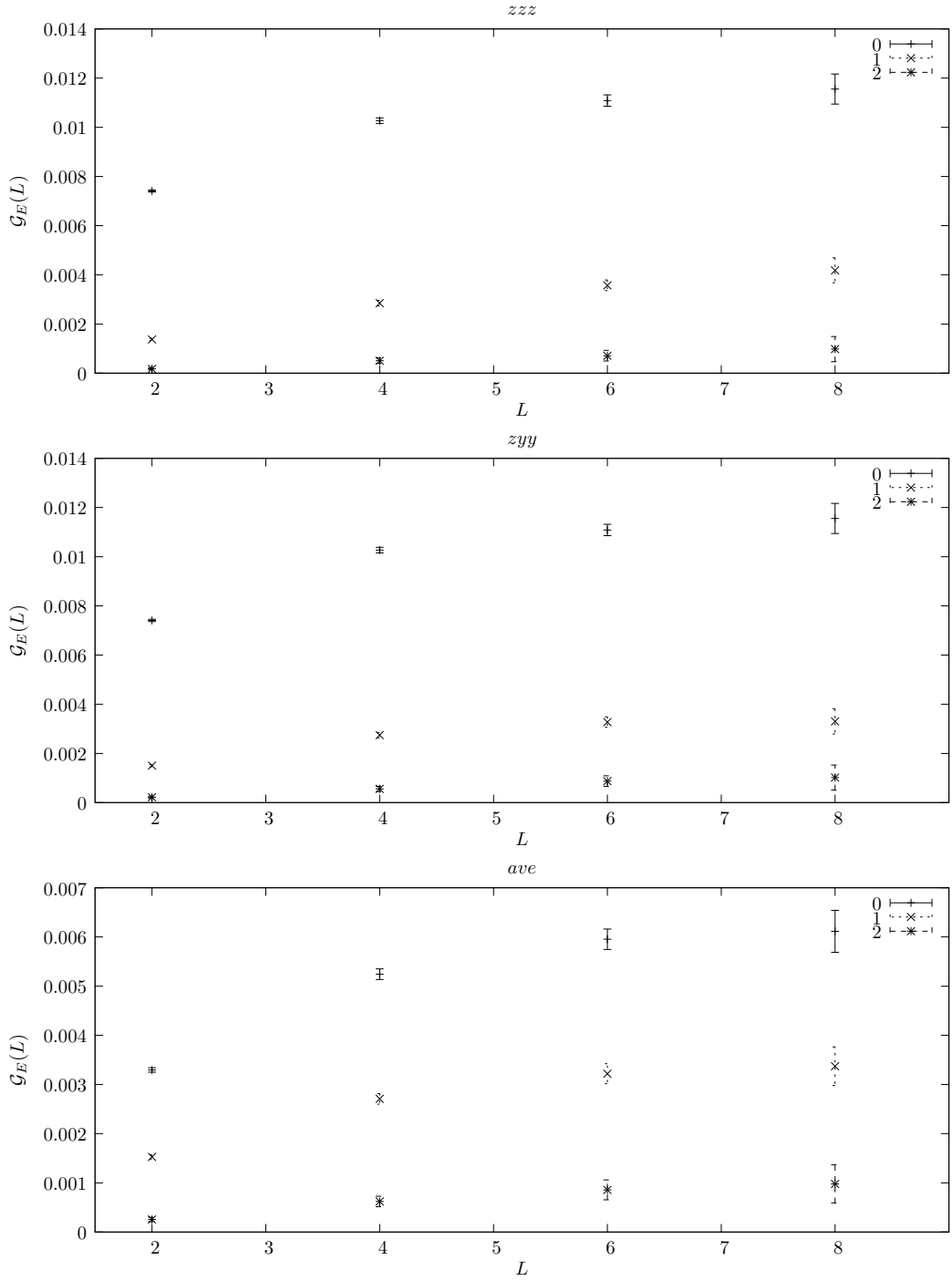}}
    \vspace{1.5cm}
    \caption{Dependence of ${\cal G}_E$ on the length $L_1=L_2=L$ (in lattice
    units) of the loops at $\theta=90^{\circ}$ for $d=0,1,2$.}
    \label{fig:Tdep}
  \end{center}
\end{figure}

\begin{figure}[h]
  \begin{center}
    \resizebox{\textwidth}{!}{\includegraphics{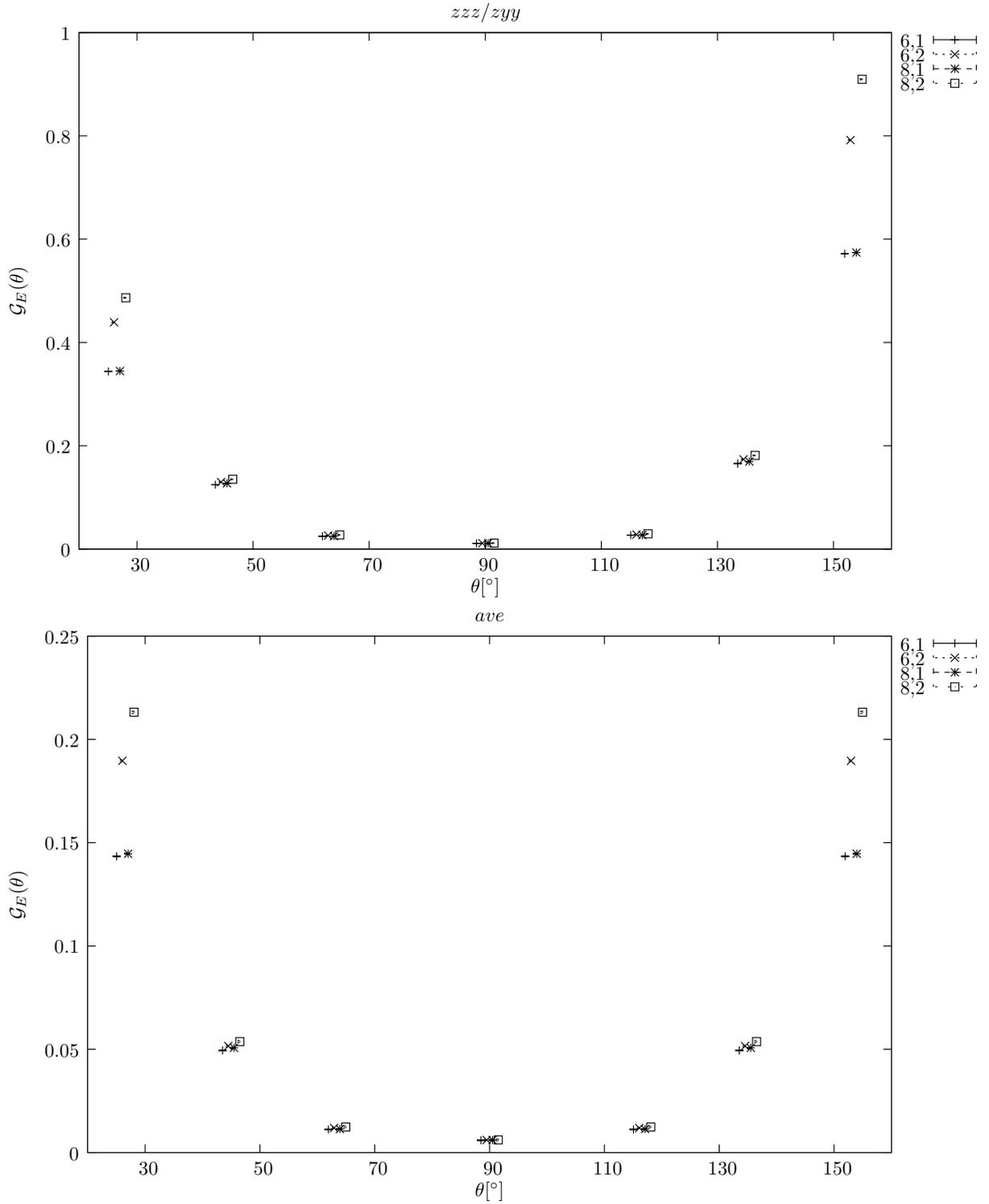}}
    \caption{Angular dependence of ${\cal G}_E$ for various lengths of the
      loops, for $d=0$. The numbers in the key refer to the on--axis loop
      length and to the off--axis loop set respectively, as explained in the
      text and in Table \ref{tab:L2}. (Different data sets have been slightly
      shifted horizontally for clarity.)}
    \label{fig:ang_gen1}
  \end{center}
\end{figure}

\begin{figure}[h]
  \begin{center}
    \resizebox{\textwidth}{!}{\includegraphics{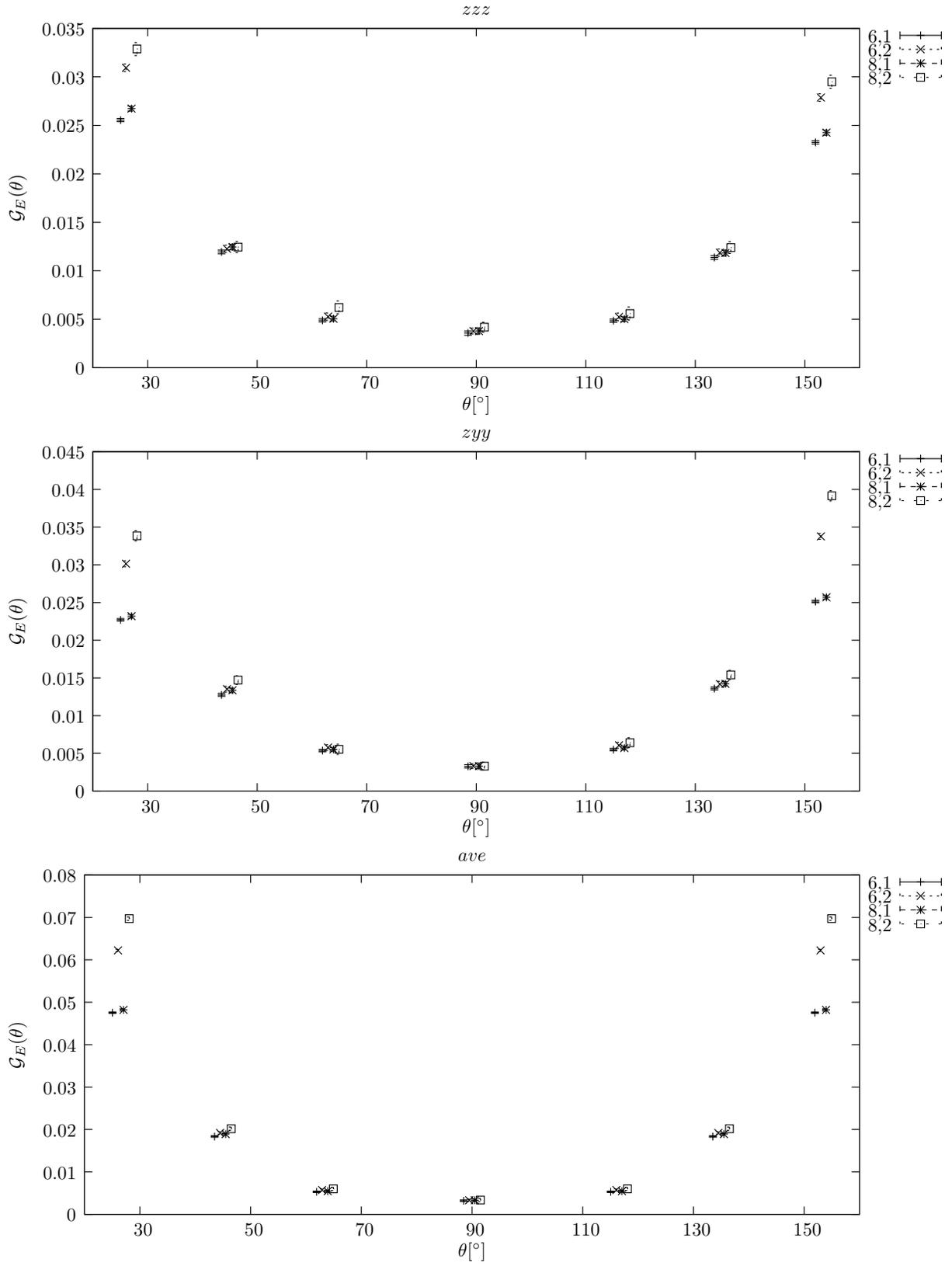}}
    \caption{Angular dependence of ${\cal G}_E$ for various lengths of the
      loops, for $d=1$. The notation is the same as in Fig.~\ref{fig:ang_gen1}.}
    \label{fig:ang_gen2}
  \end{center}
\end{figure}

\begin{figure}[h]
  \begin{center}
    \resizebox{\textwidth}{!}{\includegraphics{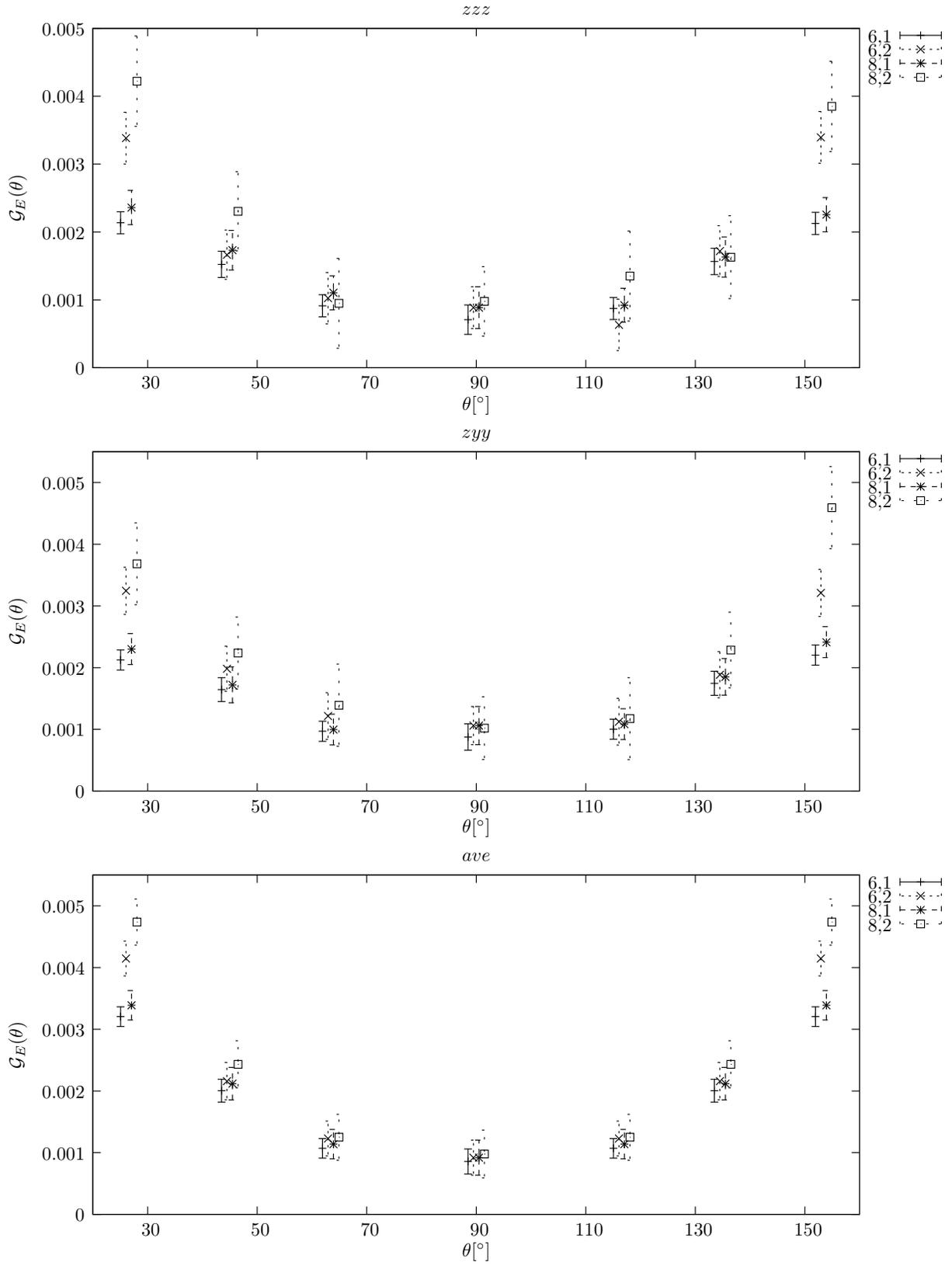}}
    \caption{Angular dependence of ${\cal G}_E$ for various lengths of the
      loops, for $d=2$.  The notation is the same as in Fig.~\ref{fig:ang_gen1}.}
    \label{fig:ang_gen3}
  \end{center}
\end{figure}

As explained in Section 2, we are interested in the $T\to\infty$
limit and so we have to somehow perform it on the lattice. In
practice, we have to look for a {\it plateau} of the correlation
function plotted against the loop lengths $L_1$ and $L_2$: in
Fig.~\ref{fig:Tdep} we show the dependence of the correlator on the length 
$L_1=L_2=L$ of the loops at $\theta=90^{\circ}$. Of course, on
a $16^4$ lattice it is difficult to have a sufficiently long loop 
while at the same time avoiding finite size effects and at best
we can push the calculation up to $L=8$; nevertheless, a {\it plateau} seems 
to have been practically reached at about $L=L_{\rm pl} \simeq
6\textendash 8$. As
$\theta$ varies from $90^{\circ}$ towards $0^{\circ}$ or $180^{\circ}$, we
expect $L_{\rm pl}$ to grow\footnote{Indeed, $L_{\rm pl}$ blows up at 
$0^{\circ},180^{\circ}$ due to the relation between the correlation function
and the static dipole--dipole potential to be discussed below, see
Eq.~(\ref{eq:Pot}).}; however, the plots in
Figs.~\ref{fig:ang_gen1}--\ref{fig:ang_gen3} show that the correlation  
function is already quite stable against variations of the loop lengths at
$L_1,L_2\simeq 8$ (at least for $\theta$ not too close to $0^{\circ}$ or
$180^{\circ}$) and so we can take the data for the largest loops available as a 
reasonable approximation of ${\cal 
C}_L$, defined as the asymptotic value of ${\cal G}_L$ as $L_1,L_2\to\infty$.
We estimate the uncertainty in ${\cal C}_L$ due to this
approximation from the variation of ${\cal G}_L$ with the lenghts $L_1,L_2$ as
(using the notation introduced in Table \ref{tab:L2}) 
\bea
  (\delta{\cal C}_L)_{syst}
  = \dfrac{1}{2}\big\{|{\cal G}_L(L_1=8,{\rm set}=2) &-& {\cal G}_L(L_1=8,{\rm set}=1)| \\
  + |{\cal G}_L(L_1=8,{\rm set}=2) &-& {\cal G}_L(L_1=6,{\rm set}=2)|
    \big\}. \nonumber 
\eea
The errors shown in Figs.~\ref{fig:ang_gen1}--\ref{fig:ang_gen3}
are the statistical ones only; in all the other figures we show the
total error obtained adding (in quadrature) the statistical and the
``systematic'' errors defined above.\\
As already noticed above, the data are less stable for $\theta$ near $0^{\circ}$
and $180^{\circ}$: this is a consequence of the relation between the
correlation function ${\cal G}_E$ and the static dipole--dipole
potential $V_{12}$~\cite{Pot},
\be
\label{eq:Pot}
{\cal G}_E(\theta = 0;T;\vec{z}_\perp;\vec{R}_{1\perp},\vec{R}_{2\perp})
\mathop\simeq_{T \to \infty} \exp \left[
-2T~ V_{12}(\vec{z}_\perp,\vec{R}_{1\perp},\vec{R}_{2\perp}) \right] -1
\ee
(note that we have set again $T_1=T_2=T$),
from which we expect ${\cal G}_E$ to diverge at $\theta=0^{\circ}$;
by virtue of the {\it crossing--symmetry relations} (\ref{crossing}), 
a similar singularity is also expected at
$\theta=180^{\circ}$~\cite{crossing}. In the following we will
consider only $\theta\neq 0^{\circ},180^{\circ}$. 

\begin{figure}
  \begin{center}
    {\includegraphics{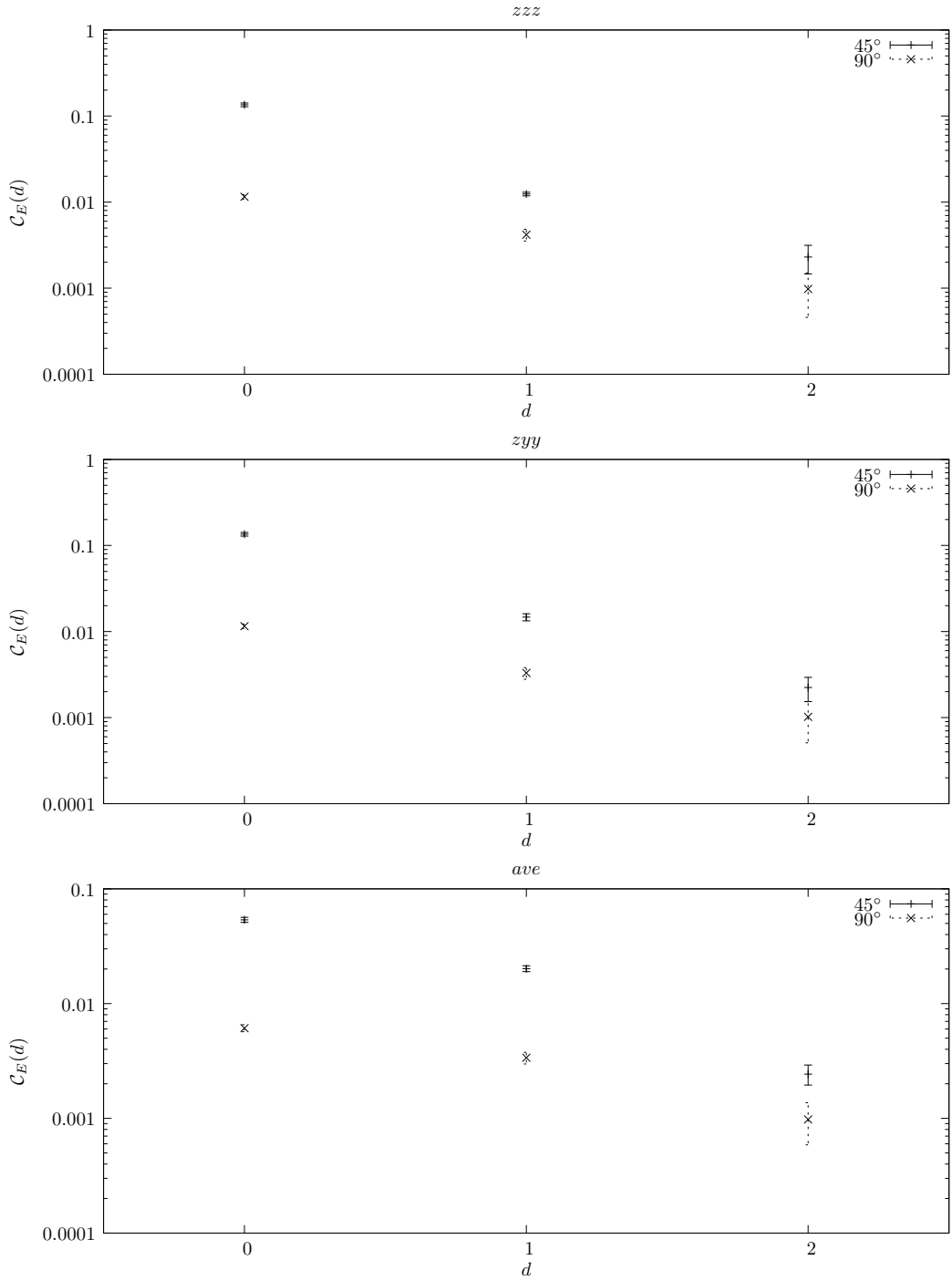}}
    \caption{Dependence of ${\cal C}_E(d)$ (on a logarithmic scale) on the
      distance $d$ (in lattice units) at $\theta=45^{\circ}$ and $\theta=90^{\circ}$ .}
    \label{fig:ddep}
  \end{center}
\end{figure}

We have considered the values $d=0,1,2$ for the distance between the
centers of the loops. As expected (see the previous Section), the
correlation functions vanish rapidly as $d$ increases, as can be seen
in Fig.~\ref{fig:ddep}, thus making the calculation with our simple ``brute force''
approach very difficult at larger distances.
\begin{table}[t]
  \centering
\begin{tabular*}{0.9\textwidth}{@{\extracolsep{\fill}}cc|*{7}{c}}

    & $\theta$ & $26.565^{\circ}$ &
    {$45^{\circ}$} &
    {$63.435^{\circ}$} & 
    {$90^{\circ}$} &
    {$116.565^{\circ}$} &
    {$135^{\circ}$} &
    {$153.435^{\circ}$}\\
    \hline
    &&&&&&&&
    \\
    $\vec{l}_{2\parallel}$ & set 1 & $(4,2)$ & $(4,4)$ & $(2,4)$ & $(0,6)$
    & $(-2,4)$ & $(-4,4)$ & $(-4,2)$ \\%
    & set 2 & $(8,4)$ & $(6,6)$ & $(4,8)$ & $(0,8)$ & $(-4,8)$ & $(-6,6)$ & $(-8,4)$ \\%
    \hline
    &&&&&&&&
    \\
    $L_2$ & set 1 & $2\sqrt{5}$ & $4\sqrt{2}$ & $2\sqrt{5}$ & $6$ &
    $2\sqrt{5}$ & $4\sqrt{2}$ & $2\sqrt{5}$\\
    & set 2 & $4\sqrt{5}$ & $6\sqrt{2}$ & $4\sqrt{5}$ & $8$ &
    $4\sqrt{5}$ & $6\sqrt{2}$ & $4\sqrt{5}$\\
\hline
\end{tabular*}
  \caption{Longitudinal vector $\vec{l}_{2\parallel}$ and length $L_2$
    for the various angles and for the two sets of off--axis loops
    $\widetilde{\cal W}_{L2}$.}
  \label{tab:L2}
\end{table}

From now on we will discuss the issue of the angular dependence of the
correlation function. 
As already pointed out in the Introduction, numerical simulations of
LGT can provide the Euclidean correlation function only for a finite
set of $\theta$--values, and so its analytic properties cannot be
directly attained; nevertheless, they are first--principles
calculations that give us (inside the errors) the true QCD expectation
for this quantity. Approximate analytic calculations of this same
function then have to be compared with the lattice 
data, in order to test the goodness of the approximations involved. 
The Euclidean correlation functions we are interested in have been
evaluated in the SVM~\cite{LLCM2}, in the ILM~\cite{instanton1} and
using the AdS/CFT correspondence (in the ``conformal''~\cite{JP1} and
``non--conformal'' cases~\cite{JP2}). The SVM  gives a well--defined
quantitative prediction, that can be tested numerically against our
data; in the ILM and conformal--AdS/CFT cases we have qualitative
knowledge of the functional dependence of the correlation functions on
the angle $\theta$ and we can test at least the goodness of the
functional form with a fit to the lattice data; in the
non--conformal--AdS/CFT case the explicit $\theta$--dependence is
unknown and we are unable to make any comparison with our data.

\begin{figure}
  \centering
  \resizebox{\textwidth}{!}{\includegraphics{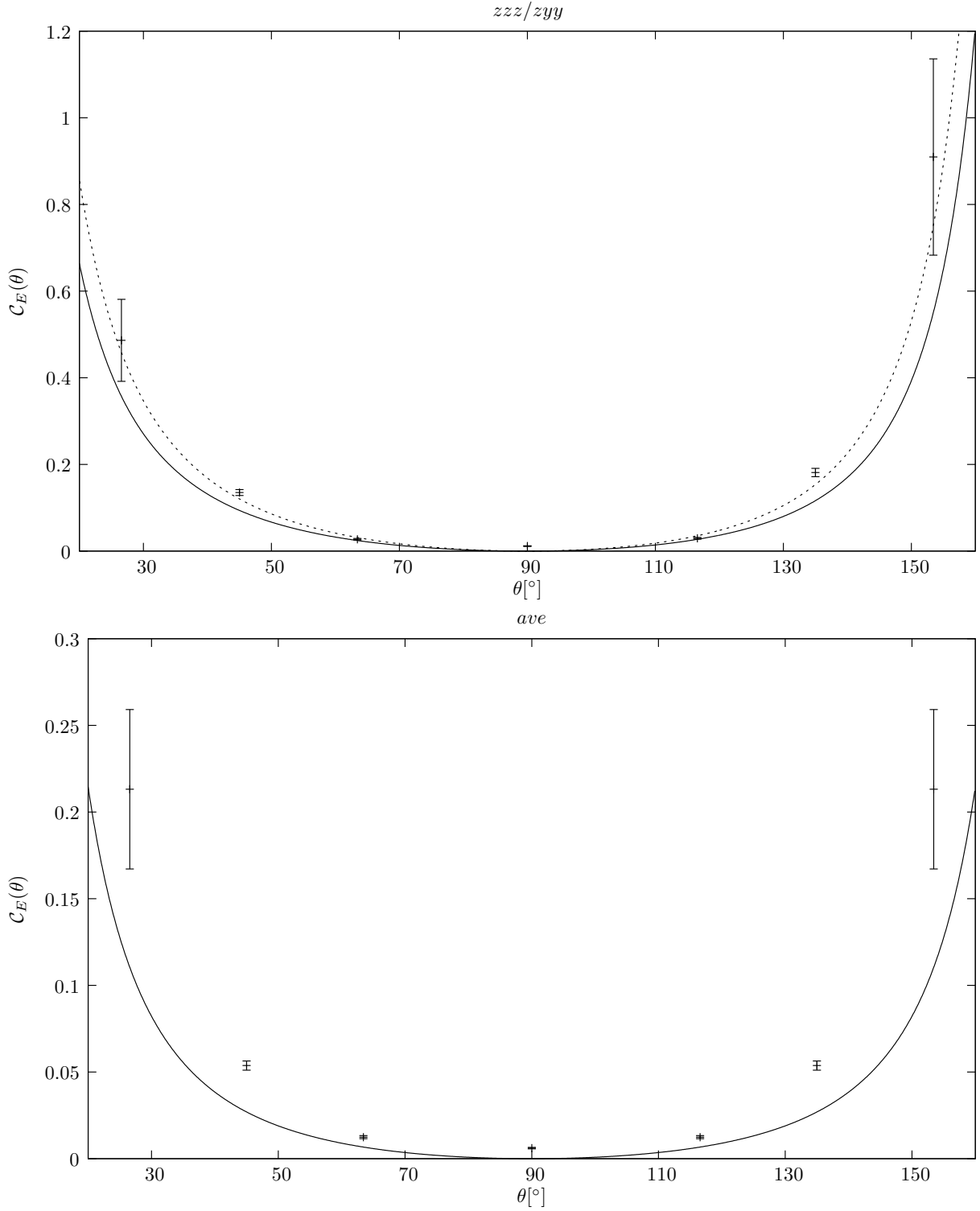}}
  \caption{Comparison of the lattice data to the SVM prediction (\ref{eq:SVM})
    with $K_{\rm SVM}$ calculated according to Ref.~\cite{LLCM2} (solid line) 
    and to the one--parameter ($K_{\rm SVM}$) best--fit (for the ``{\it zzz}''
    and ``{\it zyy}'' cases only) with the SVM expression (\ref{eq:SVM}) 
    (dotted line) at $d=0$.}
\label{fig:svm0}
\end{figure}

\begin{figure}
  \centering
  \resizebox{\textwidth}{!}{\includegraphics{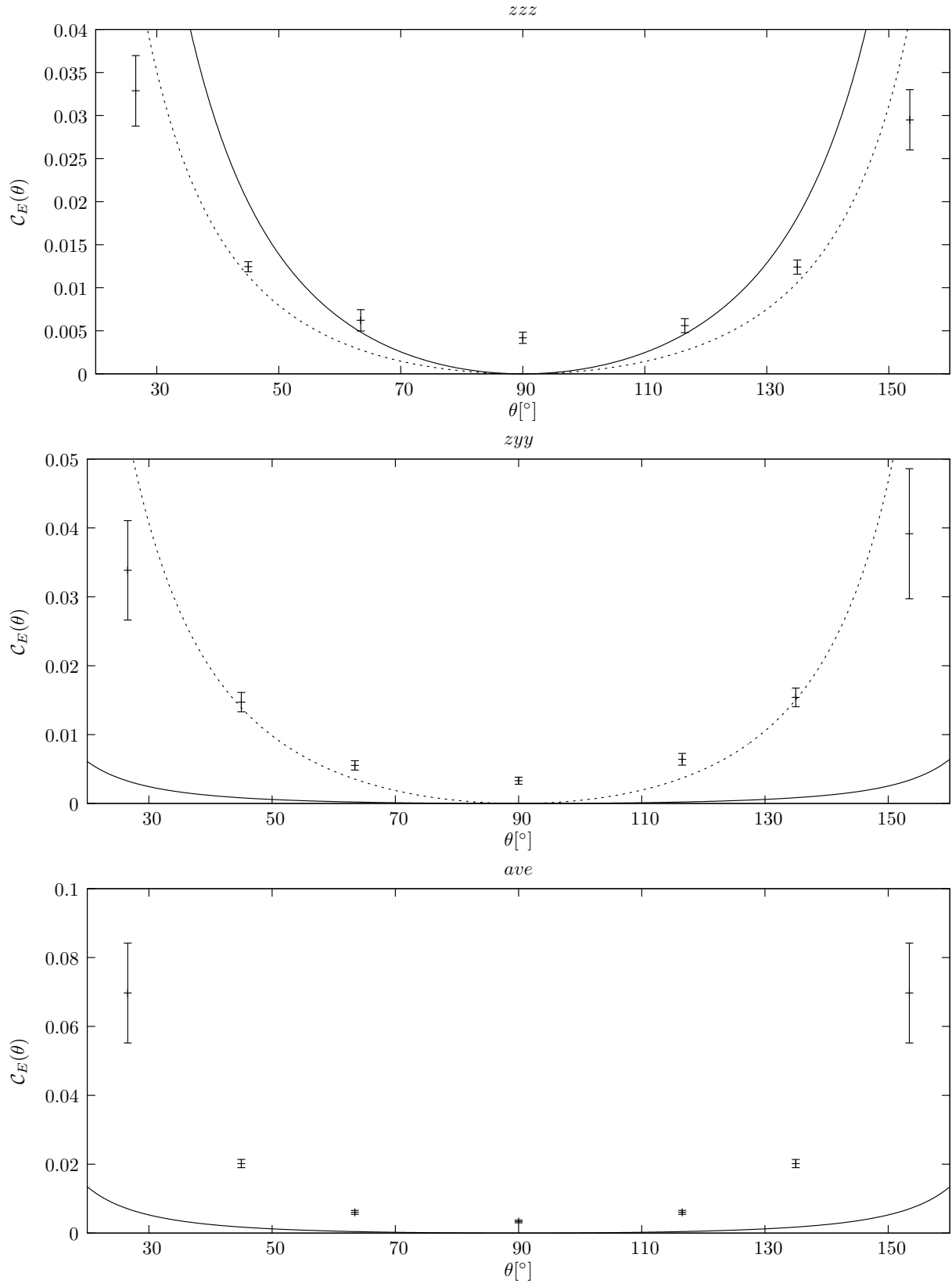}}
  \caption{The same comparison as in Fig.~\ref{fig:svm0}, but at $d=1$.}
\label{fig:svm1}
\end{figure}

\begin{figure}
  \centering
  \resizebox{\textwidth}{!}{\includegraphics{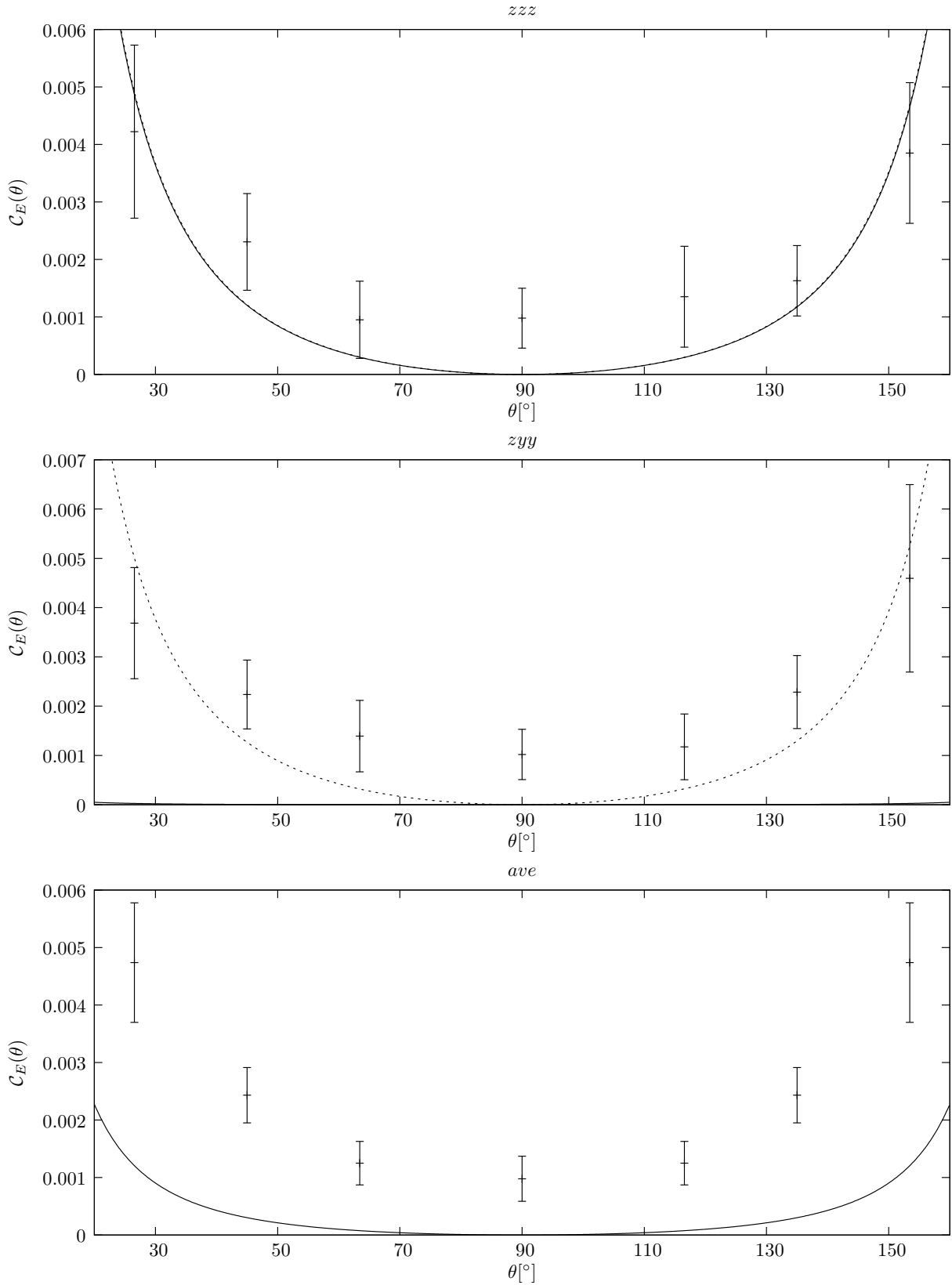}}
  \caption{The same comparison as in Fig.~\ref{fig:svm0}, but at $d=2$.}
\label{fig:svm2}
\end{figure}

In the SVM~\cite{LLCM2} the Wilson--loop correlation function
in $N_c$--colour QCD ($N_c=3$ in our case) is given by the
expression\footnote{Here and in the following formulae we 
  omit the variables other than $\theta$ on which ${\cal C}_E$
  depends.} 
\bea
 {\cal C}^{\,\rm (SVM)}_E(\theta) &=&
\left(\dfrac{N_c+1}{2N_c}\right)
\exp\left[-\left(\dfrac{N_c-1}{2N_c}\right)K_{\rm SVM}\cot\theta\right]
\nonumber 
  \label{eq:SVM}\\ & + &
\left(\dfrac{N_c-1}{2N_c}\right)
\exp\left[\left(\dfrac{N_c+1}{2N_c}\right)K_{\rm SVM}\cot\theta\right] - 1, 
\eea
where $K_{\rm SVM}$ is a function of $\vec{z}_{\perp}$, $\vec{R}_{1\perp}$
and $\vec{R}_{2\perp}$ only, whose precise expression, which we have used to
numerically evaluate the correlator (\ref{eq:SVM}) in the relevant cases, is
given in Ref.~\cite{LLCM2}. The comparison of the SVM prediction
for ${\cal C}_E$  with our data is shown in
Figs.~\ref{fig:svm0}--\ref{fig:svm2}; in the ``{\it ave}'' case the
comparison is made with ${\cal C}^{ave}$ as defined in Eq.~(\ref{eq:ave}),
with $f_1=f_2=0$ (see discussion above). Although in some cases the agreement
is quite good, at least in the shape of the curve and in the order of
magnitude, in general it is far from being satisfactory; in particular, in the
``{\it zyy}'' case for $d\neq 0$ the SVM prediction is orders of magnitude
smaller than the lattice results. One can also go the other way round, namely
try to determine $K_{\rm SVM}$ in the ``{\it zzz}'' and ``{\it zyy}'' cases
with a one--parameter best--fit to the data: the results are shown in
Figs.~\ref{fig:svm0}--\ref{fig:svm2}. In general, the difference between
the predicted and the fitted values for $K_{\rm SVM}$ is positive and in some
cases the discrepancy is larger than 20\%, which can be taken as the accuracy
level of the model parameters (see Ref.~\cite{LLCM2}). We note that for
$d=0,1$ the main contribution to the value of $K_{\rm SVM}$, evaluated using
the SVM expression given in Ref.~\cite{LLCM2} (which, we recall, consists of a
{\it perturbative} plus a {\it nonperturbative} component) comes from perturbative
effects, while at $d=2$ nonperturbative effects are equal to or greater than the
perturbative ones. We have also tried a best--fit with the following simple
functional form:
\be
  \label{eq:pert}
  {\cal  C}_E^{(\rm pert)}(\theta) = K_{\rm pert} (\cot\theta)^2,
\ee
which is exactly what one obtains in leading--order perturbation
theory~\cite{BB,Meggiolaro05,LLCM2}. Notice, however, that the coefficient
$K_{\rm pert}$ in Eq.~(\ref{eq:pert}) can also receive nonperturbative
contributions, as one can see, for example, when expanding the
exponentials in the SVM expression (\ref{eq:SVM}) to first
order\footnote{This makes sense when $K_{\rm SVM}$ is small, as it
  happens for example in the large--$N_c$ expansion, 
  where it is $K_{\rm SVM} = {\cal O}(\frac{1}{N_c})$.}.
The values of the chi--squared per
degree of freedom ($\chi^2_{\rm   d.o.f.}$) of the various fits that we have
performed are listed in Table \ref{tab:chi2}.

\begin{figure}
  \centering
  \resizebox{\textwidth}{!}{\includegraphics{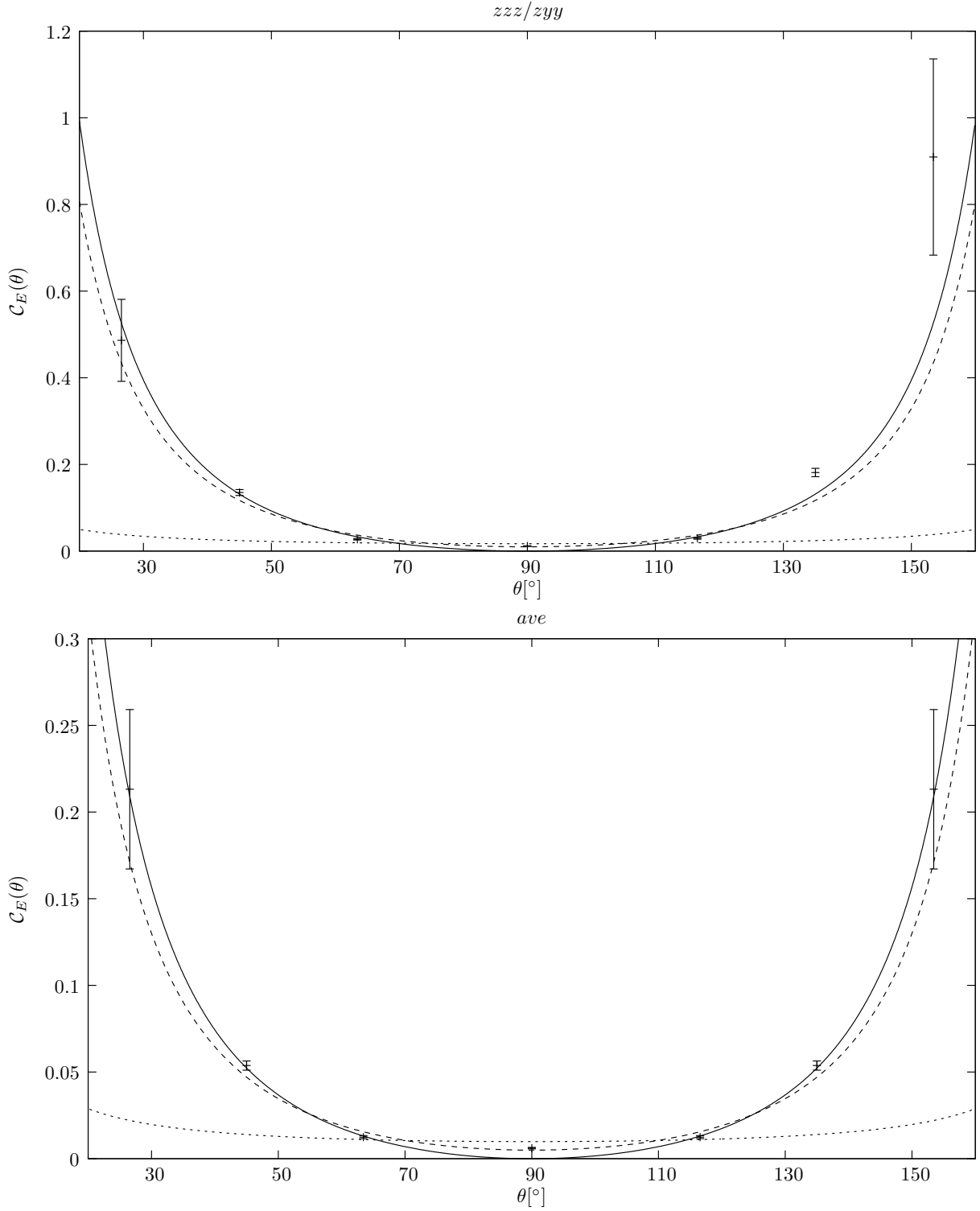}}
  \caption{Comparison of lattice data to best--fits with the
    perturbative--like expression (\ref{eq:pert}) (solid line), the
    ILM expression (\ref{eq:instanton}) (dotted line) and the ILMp
    expression (\ref{eq:instpert}) (dashed line) at $d=0$.}
  \label{fig:pertinst0}  
\end{figure}

\begin{figure}
  \centering
  \resizebox{\textwidth}{!}{\includegraphics{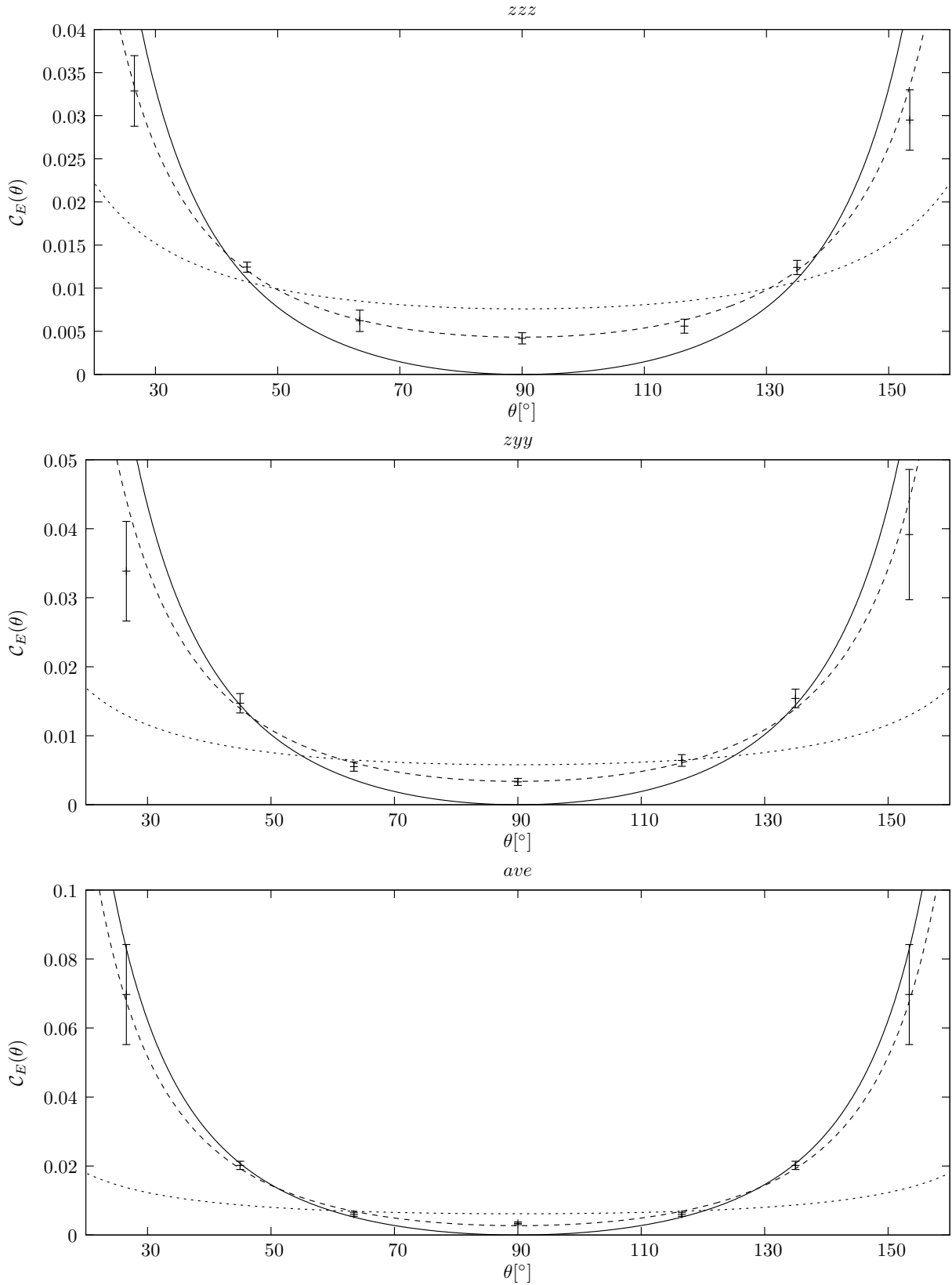}}
  \caption{The same comparison as in Fig.~\ref{fig:pertinst0}, but at $d=1$.}
  \label{fig:pertinst1}
\end{figure}

One--instanton effects in the ILM give the following analytic
expression for the correlation function~\cite{instanton1}: 
\be
\label{eq:instanton}
{\cal C}^{\,\rm (ILM)}_E(\theta) = \frac{K_{\rm ILM}}{\sin\theta}.
\ee
Lattice data are not well fitted by such a function, at least at $d=0,1$; at
$d=2$ practically all the fits we have tried are good, as one can see in Table
\ref{tab:chi2}, but we interpret this as the result of the large data errors.
We can largely improve the fits by adding a term proportional to $(\cot\theta)^2$,
\be
\label{eq:instpert}
{\cal C}^{\,\rm (ILMp)}_E(\theta) = \frac{K_{\rm ILMp}}{\sin\theta}+ K_{\rm ILMp}'(\cot\theta)^2,
\ee
which describes contributions both from two--instanton~\cite{instanton1}
and, as we have already said above, from leading--order perturbative
effects; in the following, we will refer to Eq.~(\ref{eq:instpert}) as
the ``ILMp'' expression.
Differently from the SVM case (and also from the AdS/CFT case to be
discussed below), the dependence
on $\theta$ is not affected by the average over transverse orientations (and
longitudinal--momentum fractions), so it makes sense to try to also fit the
``averaged'' data with these functional forms. The resulting best--fit
functions in the $d=0,1$  cases are plotted in
Figs.~\ref{fig:pertinst0} and \ref{fig:pertinst1}.
As the fit gives $K_{\rm ILMp}'\gg K_{\rm ILMp}^2$, it seems more
likely that $K_{\rm ILMp}'$ is dominated by perturbative effects;
while $K_{\rm ILMp}'$ is an order of magnitude larger than
$K_{\rm ILMp}$  at $d=0$, suggesting that perturbative effects are dominant,
at $d=1$ the two parameters are comparable, and at $d=2$ $K_{\rm ILMp}$
becomes larger than $K_{\rm ILMp}'$, suggesting that we are entering the
nonperturbative region.

\begin{figure}
  \centering
  \resizebox{\textwidth}{!}{\includegraphics{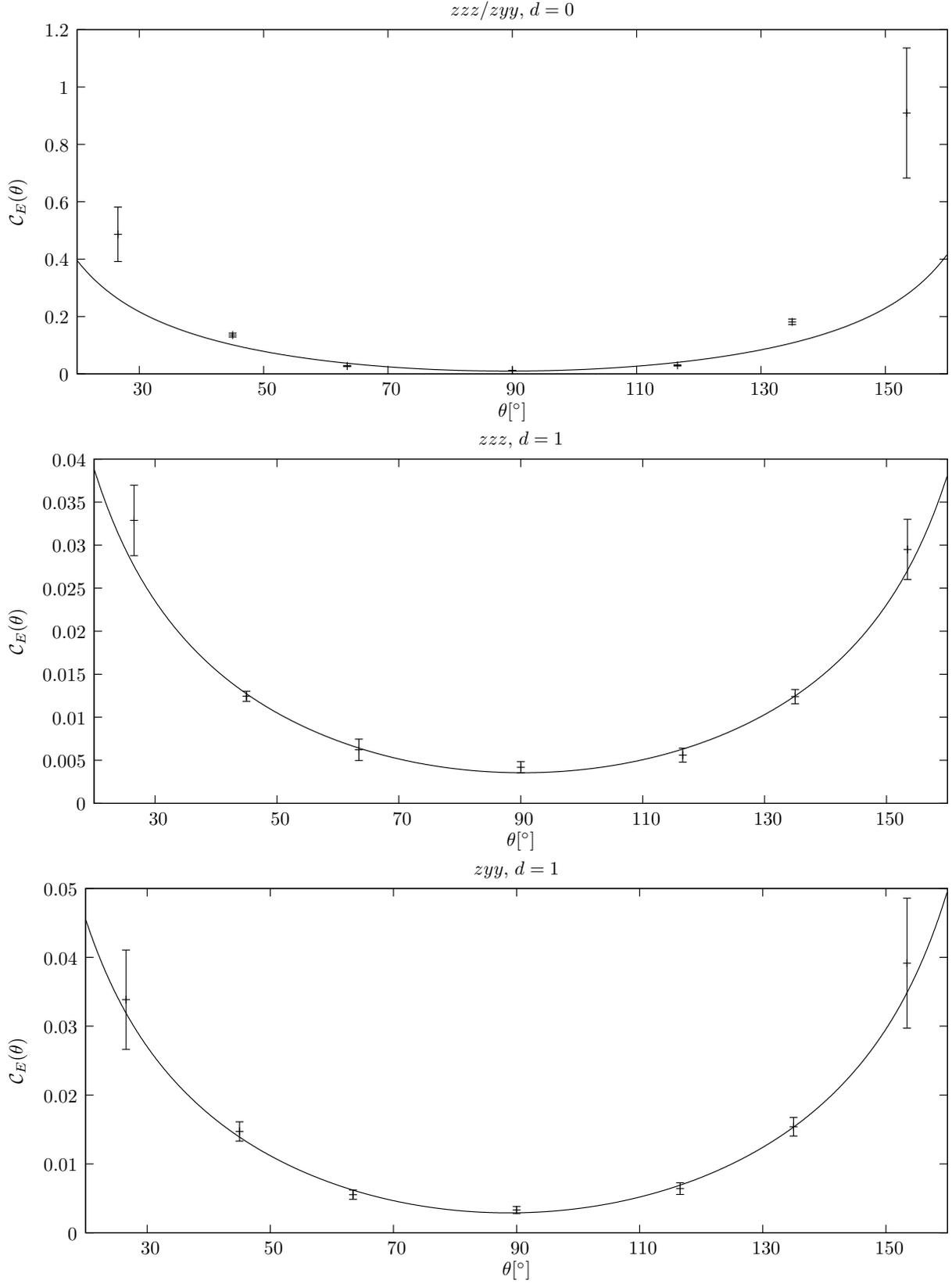}}
  \caption{Comparison of lattice data to a best--fit with the AdS/CFT
    expression (\ref{eq:AdS}) for various cases.}
    \label{fig:ads}
\end{figure}

Using the AdS/CFT correspondence, one obtains for the ${\cal N}=4$ SYM
theory at large $N_c$ and large 't Hooft coupling, and at large
distances between the loops~\cite{JP1}
\be
\label{eq:AdS}
{\cal C}^{\,\rm (AdS/CFT)}_E(\theta)
= \exp\left\{K_1\frac{1}{\sin\theta} + K_2\cot\theta +
  K_3\cos\theta\cot\theta\right\}-1.
\ee 
Although, of course, there is apparently no reason for this formula to apply
in our case, we nevertheless have tried to fit the lattice data (in the
``{\it zzz}'' and ``{\it zyy}'' cases) also with this functional form; the results
are shown in Fig.~\ref{fig:ads}. The fit is not good at $d=0$; it may seem
good at $d=1,2$, but we must remember that this is a three--parameter fit,
i.e., there are only four degrees of freedom: the ILMp best--fit, which has
only two parameters (i.e., five degrees of freedom) achieves a better or
comparable $\chi^2_{\rm d.o.f.}$ and, moreover, we have explicitly
tested some modified versions of the SVM expression (\ref{eq:SVM}) with two
and three parameters in the exponents, obtaining smaller values for
$\chi^2_{\rm d.o.f.}$.
\begin{table}[t]
  \centering
    \begin{tabular}[h]{l|cc|ccc|ccc}
      $\chi^2_{\rm d.o.f.}$ & \multicolumn{2}{c|}{$d=0$} &
      \multicolumn{3}{c|}{$d=1$} & \multicolumn{3}{c}{$d=2$}\\
      & {\it zzz}/{\it zyy} & {\it ave} & {\it zzz} & {\it zyy} &
                  {\it ave} & {\it zzz} & {\it zyy} & {\it ave} \\
                  \hline
                  \hline
                  SVM & 51 & - &  16  & 12 & - & 1.5 & 2.2 & -\\
                  pert & 53 & 34 & 16 & 13 & 13 & 1.5 & 2.2 & 4.5\\
                  ILM & 114 & 94 & 14 & 15 & 45 & 0.45 & 0.35 & 1.45\\
                  ILMp & 20 & 9.4 & 0.54 & 0.92 & 1.8 & 0.13 & 0.12 & 0.19\\
                  AdS/CFT & 40 & - & 1 & 0.63 & - & 0.14 & 0.065 & -
    \end{tabular}
  \caption{Chi--squared per degree of freedom for a best--fit with the
  indicated function.}
  \label{tab:chi2}
\end{table}

As we have said in the Introduction, the main motivation in studying soft
high--energy scattering is that it can lead to a resolution of the total cross
section puzzle, so it is worth discussing what the various models have to say
on this point. Using Eq.~(\ref{eq:WLoptic0}) it is easy to see that the SVM,
the ILM and the lowest order perturbative expressions give constant cross
sections at high energy\footnote{Actually, the ILM expression
  (\ref{eq:instanton}) for the correlation function (where we have to
  recall that $K_{\rm ILM}$ is a {\it real} function) results in an
  exactly {\it zero} total cross section, when inserted in Eq.~(\ref{eq:WLoptic0}).},
as in these cases the high--energy limit can be carried over under the
integral sign, so that the knowledge of the $\theta$--dependence of
the correlation function is sufficient to completely determine the
high--energy behaviour of total cross sections. However, it may happen
that the integrand in Eq.~(\ref{eq:WLoptic0}) has no definite
asymptotic behaviour at large $s$, as it happens, e.g., for the AdS/CFT
expression (\ref{eq:AdS}), so that the remaining integrations
have to be carried out before taking $s\to\infty$. Anyway, as the
experimentally observed universality suggests that the high--energy behaviour
is not affected by the details of the hadron wave functions (and thus by the
detailed dependence of the correlation function on $\vec{R}_{i\perp}$ and
$f_i$), we would expect that the high--energy asymptotics could be read off
after integrating only over the distance $|\vec{z}_{\perp}|$ between the
loops and this would require the detailed knowledge of the dependence of the 
correlators on $|\vec{z}_{\perp}|$. As already noticed above, however, lattice
data show that the relevant correlation functions vanish rapidly with the
distance: due to our limited numerical knowledge we postpone the detailed
discussion of the dependence on the distance and related issues to subsequent
works.

\begin{figure}
  \centering
  \resizebox{\textwidth}{!}{\includegraphics{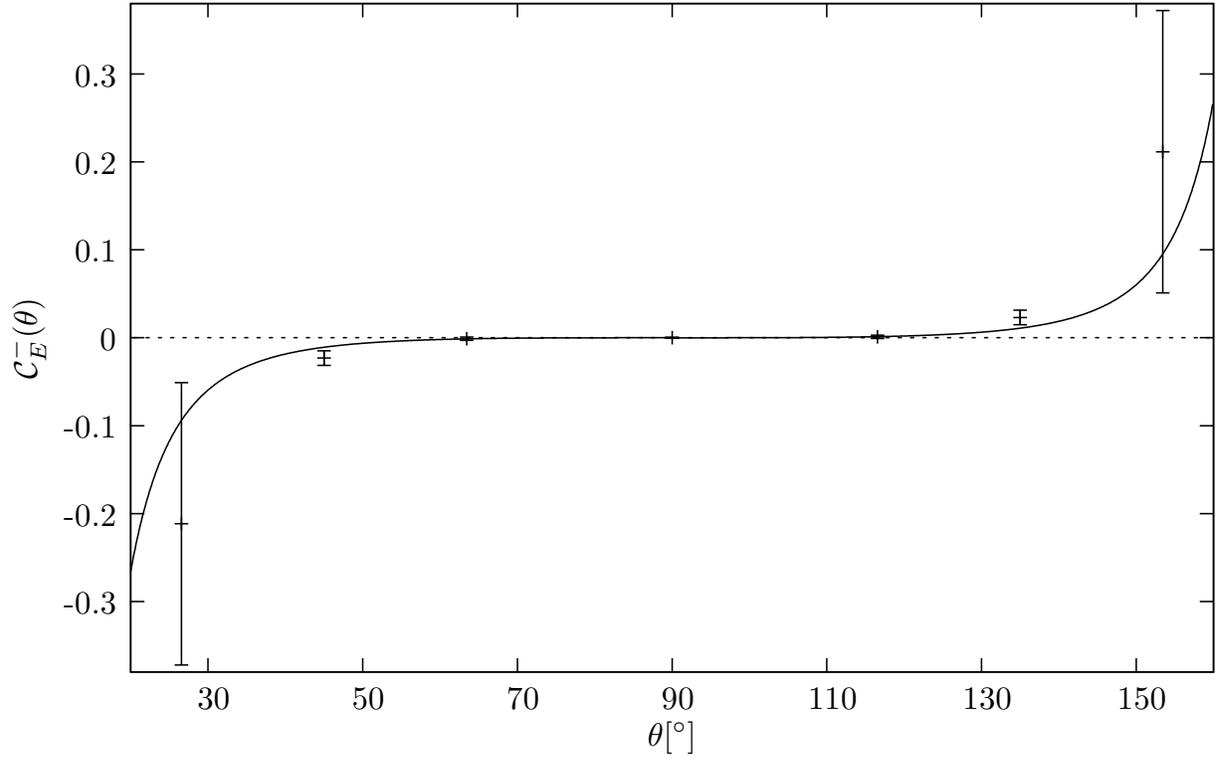}}
  \caption{The {\it crossing--antisymmetric} component ${\cal C}_E^-$, as
    defined in Eq.~(\ref{eq:antisym}), for the ``{\it zzz}/{\it zyy}'' case at $d=0$ and the
    corresponding prediction using the SVM expression (\ref{eq:SVM}) (solid line).} 
  \label{fig:odderon}
\end{figure}

As a final and important remark, we note that our data show a clear signal of
$C$--odd contributions  in dipole--dipole scattering. As shown in
Ref.~\cite{crossing} (and briefly recalled in Section 2) the
loop--antiloop correlator at angle $\theta$ in the Euclidean theory
(or at hyperbolic angle $\chi$ in the Minkowskian theory) can be
derived from the corresponding loop--loop correlator by the
substitution $\theta\to\pi -\theta$ (or $\chi\to i\pi -\chi$ in the
Minkowskian theory). Because of this {\it crossing--symmetry relation}, it
is natural (see Ref.~\cite{Meggiolaro07}) to decompose the Euclidean
correlation function ${\cal C}_E(\theta)$ as a sum of a {\it
  crossing--symmetric} function ${\cal C}_E^+(\theta)$ and a {\it
  crossing--antisymmetric} function ${\cal C}_E^-(\theta)$, 
\be
\label{eq:antisym}
{\cal C}_E(\theta)={\cal C}_E^+(\theta)+ {\cal C}_E^-(\theta), \qquad
{\cal C}_E^{\pm}(\theta)\equiv \dfrac{{\cal C}_E(\theta)\pm {\cal C}_E(\pi
  -\theta)}{2}. 
\ee
Upon analytic continuation from the Euclidean to the
Minkowskian theory, using \linebreak 
Eq.~(\ref{crossingC}), one can show that
they are related respectively to {\it pomeron} (i.e., $C=+1$) and
{\it odderon} (i.e., $C=-1$) exchanges in the dipole--dipole
scattering amplitude. A small but non--zero {\it
  crossing--antisymmetric} component ${\cal C}_E^-$ is present in our data, thus
signalling the presence of {\it odderon} contributions to 
the loop--loop correlation functions and in turn to the dipole--dipole
scattering amplitudes. (Such contributions are absent in meson--meson
elastic scattering~\cite{DR,Meggiolaro07}, since in this case the
relevant correlation function $C_E^{ave}$ is automatically {\it
  crossing--symmetric}, as already noticed at the end of Section 2). 
In Fig.~\ref{fig:odderon} we show the {\it crossing--antisymmetric} part
${\cal C}_E^-$ of the ``{\it zzz}/{\it zyy}''
data at $d=0$, together with the corresponding SVM prediction. 

\newsection{Conclusions and outlook}

In this paper the problem of high--energy hadron--hadron (dipole--dipole)
scattering has been addressed (for the first time) from the point of view of
lattice QCD. We have performed Monte Carlo numerical computations of Euclidean
Wilson--loop correlation functions in $SU(3)$ pure--gauge LGT. The energy
dependence of soft scattering amplitudes at high energies is encoded in the
dependence of these correlation functions on the relative angle $\theta$
between the loops and can be reconstructed after the analytic continuation
$\theta\to -i\log s/m^2$. 

In this paper we have focused on the study of the angular dependence of the
Euclidean correlation function. An interesting and important feature that we
have found is the presence of an asymmetry with respect to $\theta=\pi/2$ in
the plot of the Euclidean correlation function against the relative angle:
this (upon analytic continuation from Euclidean to Minkowskian theory) is a
signal of the presence of $C$--odd contributions in dipole--dipole 
scattering, i.e., a signal of {\it odderon} exchange between the
dipoles. Even though these $C$--odd contributions are averaged to zero
in meson--meson scattering (at least in our model, as long as
the squared meson wave functions satisfy some reasonable symmetry
properties in their dependence on the dipole orientations and on the
longitudinal--momentum fractions), they might play a non--trivial role
in more general hadron--hadron processes in which baryons and
antibaryons are also involved.

Although we cannot determine exactly the angular dependence from a
finite set of numerical values, we can nevertheless compare the prediction of
any given model with the lattice data by direct numerical comparison, if the
model is quantitative, or by testing the given functional form with a
best--fit to the data, if at least the $\theta$--dependence is known; in this
way we can discriminate between various proposals and thus check the goodness
of the approximations involved in the specific models. 
[One can of course also try to fit the data with some given arbitrary
functions and then look at the results of the best--fits, but in accepting or
rejecting a given function one must also take into account physical arguments,
as two fitting functions can differ numerically by a small amount and
nevertheless have different analytic structures that can  result in completely
different (and sometimes physically unacceptable!) high--energy behaviours
after analytic continuation to Minkowski space--time.]

The comparison of our data with the existing analytic calculations is not,
generally speaking, fully satisfactory. 
\begin{itemize}
\item The SVM prediction (\ref{eq:SVM}) agrees with our lattice data in a few
  cases, at least in the shape and in the order of magnitude, but, in general,
  it is far from being satisfactory. More or less the same conclusion is
  reached if one instead performs a one--parameter best--fit with the given
  expression. This suggests that corrections to the SVM are in order, which
  could be relevant when deriving the high--energy behaviour of the scattering
  amplitudes, upon analytic continuation to Minkowski space--time.
\item We have then tried best--fits with a simple perturbative--like
  expression (\ref{eq:pert}) and with the ILM expression
  (\ref{eq:instanton}). The results are again not satisfactory. In particular,
  the ILM expression seems to be strongly disfavoured at $d=0$, while at $d=2$
  it looks better than the SVM and perturbative--like expressions. (This
  suggests that while at $d=0$ the perturbative effects are dominant, at $d=2$
  nonperturbative effects are surely relevant and cannot be neglected.) By
  combining the two previous expressions into the ILMp expression
  (\ref{eq:instpert}), largely improved best--fits have been obtained. It
  would be interesting to also have a quantitative numerical prediction
  from the ILM and see how it compares with the lattice data.
\item Finally, we have tried a best--fit with the AdS/CFT expression
  (\ref{eq:AdS}), although there is no reason for this formula to apply in
  our case, since it was originally derived for the ${\cal N}=4$ SYM theory at
  strong coupling and large impact parameter. Taking into account that this
  is a three--parameter best--fit, even this one is not satisfactory,
  especially at $d=0$. Best--fits with QCD--inspired expressions with only two
  parameters, like, e.g., the ILMp expression (\ref{eq:instpert}) [or some
  appropriate modification of the SVM expression (\ref{eq:SVM})] give smaller
  $\chi^2_{\rm d.o.f.}$.
\end{itemize}

Although the AdS/CFT expression (\ref{eq:AdS}), as said above, is not
expected to describe real QCD, it nevertheless shows how a
non--trivial high--energy behaviour could emerge  from a simple
analytic dependence on the angle $\theta$. However, in this case,
after the analytic continuation into Minkowski space--time, it is not
possible to pass to the high--energy limit under the integral sign, as
the integrand is an oscillating function of the energy, and one should
carry over the remaining integrals first. As we have already pointed
out, the integration over the distance between the loops should be the
relevant one and, depending on the detailed form of the various
coefficient functions, a variety of behaviours could emerge. 
It seems then worth investigating further the dependence of the
correlation functions on the relative distance between the loops, as
well as on the dependence on the relative angle, as they could combine in a
non--trivial way to determine the high--energy behaviour of meson--meson
total cross sections. Moreover, as already recalled in the previous Section
[see Eq.~(\ref{eq:Pot})], the study of the transverse--distance dependence of
the Euclidean correlation function ${\cal G}_E$ at $\theta=0,\pi$ would allow
to determine the static dipole--dipole potential. These and other related
issues will be addressed in future works.

In conclusion, the fact that the existing models are not able to fully explain
the lattice data, which inside the errors represent the true QCD expectation,
is a motivation to further investigate the problem of soft high--energy
scattering, both on the numerical and on the analytical side. We hope that the
interplay of numerical simulations in LGT and analytic nonperturbative
calculations can lead to a deeper understanding of the long--standing problem
of high--energy scattering in strong interactions.

%

\newpage



\newpage

\noindent
\begin{center}
{\large\bf FIGURE CAPTIONS}
\end{center}
\vskip 0.5 cm
\noindent

\begin{itemize}
\item[\bf Fig. 1]{ The space--time configuration of the two Wilson loops
    ${\cal W}_1$ and ${\cal W}_2$ entering 
    in the expression for the dipole--dipole elastic scattering amplitude in the
    high--energy limit.}
\item[\bf Fig. 2]{ The analyticity domains of the function $\overline {\cal
      G}_E$ in the complex variable $\theta $ and of the function $\overline
    {\cal G}_M$ in the complex variable $\chi $.} 
\item[\bf Fig. 3]{ The minimal--distance prescription for a line with slope
    $\qopname \relax o{tan}\theta = 1/2$.} 
\item[\bf Fig. 4]{ The relevant Wilson--loop
    configuration. Using the $O(4)$ invariance of the Euclidean theory we have put
$p_{1E}$ parallel to the $x_{E1}$ axis.}
\item[\bf Fig. 5]{ Loop configuration in the transverse plane. In the ``{\it
      ave}'' case the link orientation is not shown as it is averaged over.} 
\item[\bf Fig. 6]{ Dependence of ${\cal G}_E$ on the length $L_1=L_2=L$ (in
    lattice units) of the loops at $\theta=90^{\circ}$ for $d=0,1,2$.}
\item[\bf Fig. 7]{ Angular dependence of ${\cal G}_E$ for various lengths of
    the loops, for $d=0$. The numbers in the key refer to the on--axis loop
    length and to the off--axis loop set respectively, as explained in the
    text and in Table 1\hbox {}. (Different data sets have been slightly
    shifted horizontally for clarity.)} 
\item[\bf Fig. 8]{ Angular dependence of ${\cal G}_E$ for various lengths of
    the loops, for $d=1$. The notation is the same as in
    Fig.~\ref{fig:ang_gen1}.}  
\item[\bf Fig. 9]{ Angular dependence of ${\cal G}_E$ for various lengths of
    the loops, for $d=2$. The notation is the same as in
    Fig.~\ref{fig:ang_gen1}.}  
\item[\bf Fig. 10]{ Dependence of ${\cal C}_E(d)$ (on a
    logarithmic scale) on the distance $d$ (in lattice units) at $\theta
    =45^{\circ }$ and $\theta =90^{\circ }$.} 
\item[\bf Fig. 11]{ Comparison of the lattice data to the SVM
    prediction (\ref{eq:SVM}) with $K_{\rm SVM}$ calculated according
    to Ref.~\cite {LLCM2} (solid line) and to the one--parameter
    ($K_{\rm SVM}$) best--fit (for the ``{\it zzz}'' and ``{\it zyy}''
    cases only) with the SVM expression (\ref{eq:SVM}) (dotted line)
    at $d=0$.} 
\item[\bf Fig. 12]{ The same comparison as in Fig.~\ref{fig:svm0},
    but at $d=1$.} 
\item[\bf Fig. 13]{ The same comparison as in Fig.~\ref{fig:svm0},
    but at $d=2$.} 
\item[\bf Fig. 14]{ Comparison of lattice data to best--fits with the
    perturbative--like expression (\ref{eq:pert}) (solid line), the ILM expression
    (\ref{eq:instanton}) (dotted line) and the ILMp expression
    (\ref{eq:instpert}) (dashed line) at $d=0$.} 
\item[\bf Fig. 15]{ The same comparison as in Fig.~\ref{fig:pertinst0},
    but at $d=1$.} 
\item[\bf Fig. 16]{ Comparison of lattice data to a best--fit with the AdS/CFT
    expression (\ref{eq:AdS}) for various cases.}
\item[\bf Fig. 17]{ The {\it crossing--antisymmetric} component ${\cal
      C}_E^-$, as defined in Eq.~(\ref{eq:antisym}), for the ``{\it
      zzz}/{\it zyy}'' case at $d=0$ and the corresponding prediction using
    the SVM expression (\ref{eq:SVM}) (solid line).}
\end{itemize}


\newpage

\pagestyle{empty}

\clearpage

\clearpage

\clearpage

\clearpage

\clearpage

\clearpage

\clearpage

\clearpage

\clearpage

\clearpage

\clearpage

\clearpage

\clearpage

\clearpage

\clearpage

\clearpage

\end{document}